\documentclass[sigconf]{acmart}
\AtBeginDocument{%
  \providecommand\BibTeX{{%
    \normalfont B\kern-0.5em{\scshape i\kern-0.25em b}\kern-0.8em\TeX}}}

\usepackage{tikz}
\usepackage{amsmath}

\usepackage{color}
\usepackage{amsfonts}
\usepackage{amsthm}
\usepackage{xspace}
\usepackage{dsfont}
\usepackage{graphicx}
\usepackage{caption}
\usepackage{subcaption}
\usepackage{multirow}
\usepackage{url}
\usepackage{bbm}
\usepackage{booktabs}
\usepackage{enumitem}

\newif\ifarxiv
\newif\ifsubmit

\submitfalse
\arxivtrue

\newcommand{\reviewer}[3]{
  \expandafter\newcommand\csname #1\endcsname[1]{
    \ifsubmit
      \textcolor{#3}{}
    \else
      \textcolor{#3}{[#2: ##1]}
    \fi
  }
}

\ifsubmit
\newcommand{\xinyun}[1]{}
\newcommand{\wenxiao}[1]{}
\newcommand{\ruoxi}[1]{}
\newcommand{\dawn}[1]{}
\else
\newcommand{\xinyun}[1]{\textcolor{red}{[Xinyun: #1]}}
\newcommand{\wenxiao}[1]{\textcolor{green}{[Wenxiao: #1]}}
\newcommand{\ruoxi}[1]{\textcolor{blue}{[Ruoxi: #1]}}
\newcommand{\dawn}[1]{\textcolor{red}{[Dawn: #1]}}
\fi

\renewcommand{\paragraph}[1]{\noindent\textbf{#1}}
\newcommand{\bo}[1]{}

\newcommand{\eat}[1]{}

\def\ours{REFIT}

\copyrightyear{2021}
\acmYear{2021}
\setcopyright{rightsretained}
\acmConference[ASIA CCS '21]{Proceedings of the 2021 ACM Asia
Conference on Computer and Communications Security}{June 7--11,
2021}{Virtual Event, Hong Kong}
\acmBooktitle{Proceedings of the 2021 ACM Asia Conference on
Computer and Communications Security (ASIA CCS '21), June 7--11,
2021, Virtual Event, Hong Kong}\acmDOI{10.1145/3433210.3453079}
\acmISBN{978-1-4503-8287-8/21/06}

\begin{document}

\fancyhead{}

\date{}

\title{REFIT: A Unified Watermark Removal Framework For Deep Learning Systems With Limited Data}

\author{Xinyun Chen}
\email{xinyun.chen@berkeley.edu}
\authornote{Equal contribution.}
\affiliation{%
  \institution{UC Berkeley}
}
\author{Wenxiao Wang}
\authornotemark[1]
\email{wangwx20@mails.tsinghua.edu.cn}
\affiliation{%
  \institution{Tsinghua University}
}
\author{Chris Bender}
\email{chrisbender@berkeley.edu}
\affiliation{%
  \institution{UC Berkeley}
}
\author{Yiming Ding}
\email{dingyiming0427@berkeley.edu}
\affiliation{%
  \institution{UC Berkeley}
}
\author{Ruoxi Jia}
\email{ruoxijia@vt.edu}
\affiliation{%
  \institution{Virginia Tech}
}
\author{Bo Li}
\email{lbo@illinois.edu}
\affiliation{%
  \institution{UIUC}
}
\author{Dawn Song}
\email{dawnsong@cs.berkeley.edu}
\affiliation{%
  \institution{UC Berkeley}
}

\renewcommand{\shortauthors}{}

\begin{abstract}
Training deep neural networks from scratch could be computationally expensive and requires a lot of training data. Recent work has explored different watermarking techniques to protect the pre-trained deep neural networks from potential copyright infringements. However, these techniques could be vulnerable to watermark removal attacks. In this work, we propose {\ours}, a unified watermark removal framework based on fine-tuning, which does not rely on the knowledge of the watermarks, and is effective against a wide range of watermarking schemes. In particular, we conduct a comprehensive study of a realistic attack scenario where the adversary has limited training data, which has not been emphasized in prior work on attacks against watermarking schemes. To effectively remove the watermarks without compromising the model functionality under this weak threat model, we propose two techniques that are incorporated into our fine-tuning framework: (1) an adaption of the elastic weight consolidation (EWC) algorithm, which is originally proposed for mitigating the catastrophic forgetting phenomenon; and (2) unlabeled data augmentation (AU), where we leverage auxiliary unlabeled data from other sources. Our extensive evaluation shows the effectiveness of {\ours} against diverse watermark embedding schemes. The experimental results demonstrate that our fine-tuning-based watermark removal attacks could pose real threats to the copyright of pre-trained models, and thus highlight the importance of further investigating the watermarking problem and proposing more robust watermark embedding schemes against the attacks.~\footnote{The code is available at~\url{https://github.com/sunblaze-ucb/REFIT}.}
\end{abstract}

	

\maketitle

\section{Introduction}
\label{sec:intro}

Deep neural networks (DNNs) have achieved great performance on a variety of application domains, and are creating tremendous business values~\cite{he2016deep,devlin2019bert}. Building these models from scratch is computationally intensive and requires a large set of high-quality annotated training samples. Various online marketplaces, such as BigML and Amazon, have emerged to allow people to buy and sell the pre-trained models. Just like other commodity software, the intellectual property (IP) embodied in DNNs needs proper protection to preserve competitive advantages of the model owner.

To protect the intellectual property of pre-trained DNNs, a widely adopted approach is \emph{watermarking}~\cite{adi2018turning,zhang2018protecting,rouhani2018deepsigns,uchida2017embedding}. A common paradigm of watermarking is to inject some specially-designed training samples, so that the model could be trained to predict in the ways specified by the owner when the watermark samples are fed into the model. In this way, a legitimate model owner can train the model with watermarks embedded, and distribute it to the model users. When he later encounters a model he suspects to be a copy of his own, he can verify the ownership by inputting the watermarks to the model and checking the model predictions. This approach has gained a lot of popularity due to the simplicity of its protocol. 

On the other hand, recent work has studied attack approaches to bypass the watermark verification process, so that the legitimate model owner is not able to claim the ownership. To achieve this goal, there are two lines of work in the literature. One line of work studies detection attacks against watermark verification~\cite{namba2019robust,hitaj2018have}. Specifically, when the input is suspected to be a watermark by the detection mechanism, the model returns a random prediction, otherwise it returns the true model prediction. Another line of work that attracts more interest is on~\emph{watermark removal attacks}, which aims at modifying the watermarked models so that they no longer predict in the ways specified by the model owner when provided with the watermark samples. In particular, most of existing work assumes the knowledge of the watermarking scheme, e.g., the approach is specifically designed for pattern-based watermarks, where each of the watermark samples is blended with the same pattern~\cite{wang2019neural,gao2019strip,chen2019deepinspect,guo2019tabor}. Although there are some latest works studying general-purpose watermark removal schemes that are agnostic to watermark embedding approaches, including pruning~\cite{zhang2018protecting,liu2018fine,namba2019robust}, distillation~\cite{yang2019effectiveness}, and fine-pruning~\cite{liu2018fine}, most of these attacks either significantly hamper the model accuracy to remove the watermarks, or are conducted with the assumption that the adversary has full access to the data used to train the watermarked model. The lack of investigation into data efficiency leaves it unclear whether such watermark removal attacks are practical in the real world.

In this paper, we propose {\ours}, a general-purpose watermark removal framework based on fine-tuning. Although previous work suggests that fine-tuning alone is not sufficient to remove the watermarks~\cite{adi2018turning,liu2018fine}, we find that by carefully designing the fine-tuning learning rate schedule, the adversary is able to remove the watermarks instead. However, when the adversary only has access to a small training set that is not comparable to the pre-training dataset, although the watermarks can still be removed, the test accuracy could also degrade. Therefore, we propose two techniques to overcome this challenge. The first technique is adapted from elastic weight consolidation (EWC)~\cite{kirkpatrick2017overcoming}, which is originally proposed to mitigate the catastrophic forgetting phenomenon, i.e., the model tends to forget the knowledge learned from old tasks when later trained on a new one~\cite{goodfellow2013empirical,kirkpatrick2017overcoming,kemker2018measuring}. The central idea behind this component is to slow down learning on model weights that are relevant to the knowledge learned for the task of interest, and keep updating other weights that were used more for memorizing watermarks.

Another technique is called unlabeled data augmentation (AU). While a large amount of labeled data could be expensive to collect, unlabeled data is much cheaper to obtain; e.g., the adversary can simply download as many images as he wants from the Internet. Therefore, the adversary could leverage inherently unbounded provisions of unlabeled samples during fine-tuning. Specifically, we propose to utilize the watermarked model to annotate the unlabeled samples, and augment the fine-tuning training data with them.

We perform a systematic study of {\ours}, where we evaluate the attack performance when varying the amount of data the adversary has access to. We focus on watermark removal of deep neural networks for image recognition in our evaluation, where existing watermarking techniques are shown to be the most effective. To demonstrate that {\ours} is designed to be agnostic to different watermarking schemes, we evaluate our watermark removal performance over a diverse set of watermark embedding approaches, and on both transfer learning and non-transfer learning. For transfer learning setting, we demonstrate that after fine-tuning with {\ours}, the resulted models consistently surpass the test performance of the pre-trained watermarked models, sometimes even when neither EWC nor AU is applied, while the watermarks are successfully removed. For non-transfer learning setting with a very limited in-distribution training set, it becomes challenging for the basic version of {\ours} to achieve a comparable test performance to the pre-trained watermarked model. With the incorporation of EWC and AU, {\ours} significantly decreases the amount of in-distribution labeled samples required for preserving the model performance while the watermarks are effectively removed. Furthermore, the unlabeled data could be drawn from a very different distribution than the data for evaluation; e.g., the label sets could barely overlap.

To summarize, we make the following contributions.

\begin{itemize}[leftmargin=*,noitemsep,topsep=0em]
\item In contrast to the previous observation of the ineffectiveness of fine-tuning-based watermark removal schemes, we demonstrate that with an appropriately designed learning rate schedule, fine-tuning is able to remove the watermarks.

\item We propose {\ours}, a watermark removal framework that is agnostic to watermark embedding schemes. In particular, to deal with the challenge of lacking in-distribution labeled fine-tuning data, we develop two techniques, i.e., an adaption of elastic weight consolidation (EWC) and augmentation of unlabeled data (AU), towards mitigating this problem from different perspectives.

\item We perform the first comprehensive study of data efficiency of watermark removal attacks, demonstrating the effectiveness of {\ours} against diverse watermarking schemes.
\end{itemize}

Our work provides the first successful demonstration of watermark removal techniques against different watermark embedding schemes when the adversary has limited data, which poses real threats to existing watermark embedding schemes. We hope that our extensive study could shed some light on the potential vulnerability of existing watermarking techniques in the real world, and encourage further investigation of designing more robust watermark embedding approaches.

\section{Model Watermarking}
\label{sec:background}

We study the watermarking problem following the formulation in~\cite{adi2018turning}. Specifically, a model owner trains a model $f_\theta$ for a task $\mathcal{T}$. Besides training on data drawn from the distribution of $\mathcal{T}$, the owner also embeds a set of watermarks $\mathcal{K}=\{(x^k, y^k)\}_{k=1}^K$ into $f_\theta$. A valid watermarking scheme should at least satisfy two properties:

\begin{itemize}[leftmargin=*,noitemsep,topsep=0em]
\item \emph{Functionality-preserving}, i.e., watermarking does not noticeably degrade the model accuracy on $\mathcal{T}$.

\item \emph{Verifiability}, i.e., $Pr(f_\theta(x^k)=y^k) \gg Pr(f'(x^k)=y^k)$ for $(x^k, y^k) \in \mathcal{K}$, where $f'$ is any other model that is not trained with the same set of watermarks. In practice, the model owner often sets a threshold $\gamma$, so that when $Pr(\hat{f}(x^k)=y^k) > \gamma$, the model $\hat{f}$ is considered to have the watermarks embedded, which could be used as an evidence to claim the ownership. We refer to $\gamma$ as the~\emph{watermark decision threshold}.
\end{itemize}

Various watermark embedding schemes have been proposed in recent years~\cite{zhang2018protecting,chen2017targeted,gu2017badnets,adi2018turning,namba2019robust,merrer2017adversarial}. The most widely studied watermarking schemes could be pattern-based techniques, which blend the same pattern into a set of images as the watermarks~\cite{chen2017targeted,gu2017badnets,adi2018turning}. Such techniques are also commonly applied for backdoor injection or Trojan attacks~\cite{liu2017trojaning,liu2017neural,shafahi2018poison}. Therefore, a long line of work has studied defense proposals against pattern-based watermarks~\cite{wang2019neural,gao2019strip,chen2019deepinspect,guo2019tabor}. Despite that these defense methods are shown to be effective against at least some types of pattern-based watermarks, they typically rely on certain assumptions of the pattern size, label distribution, etc. More importantly, it would be hard to directly apply these methods to remove other types of watermarks, which limits their generalizability. In contrast to this line of work, we study the threat model where the adversary has minimal knowledge of the pre-training process, as detailed below.

\subsection{Threat Model for Watermark Removal}

\begin{figure}[h]
    \centering
    \includegraphics[width=\linewidth]{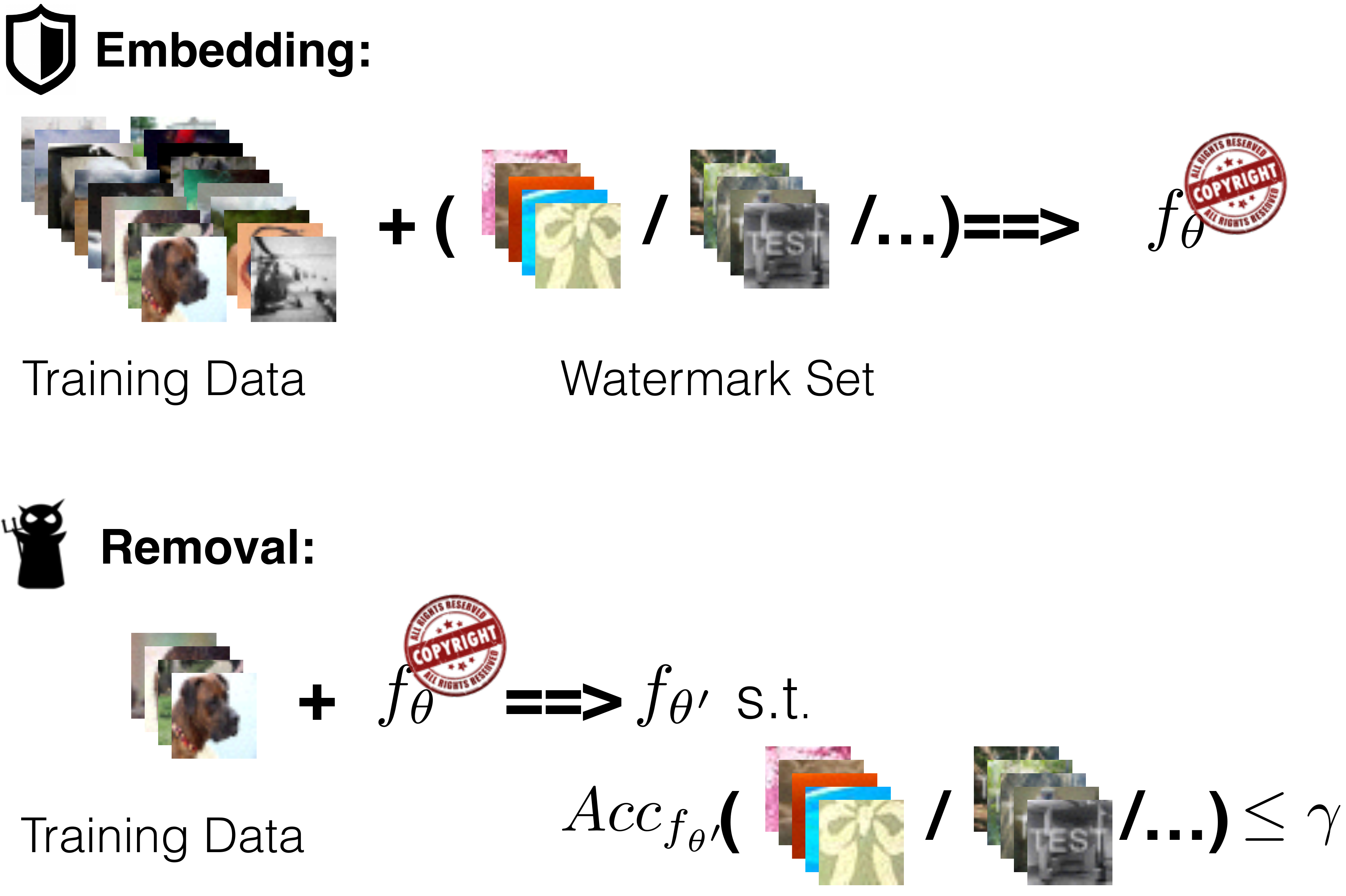}
    \caption{An overview of our setup of watermark embedding and removal, as well as the threat model. Specifically, the model owner embeds a set of watermark samples into the pre-trained model, so that these samples could be used for ownership verification. Meanwhile, the training data accessible to the adversary is too limited to train a model of good performance from scratch, which motivates the adversary to pirate a pre-trained model. To bypass the ownership verification, the adversary needs to remove the watermarks, so that the watermark accuracy does not pass the threshold $\gamma$.}
    \label{fig:threat_model}
\end{figure}

In this work, we assume the following threat model for the adversary who aims at removing the watermarks. In Figure~\ref{fig:threat_model}, we provide an overview to illustrate the setup of watermark embedding and removal, as well as the threat model.

\paragraph{No knowledge of the watermarks.} Some prior work on detecting samples generated by pattern-based techniques requires access to the entire data for pre-training, including the watermarks~\cite{tran2018spectral,chen2018detecting}. In contrast, we do not assume access to the watermarks.

\paragraph{No knowledge of the watermarking scheme.} As discussed above, most prior works demonstrating successful watermark removal rely on the assumption that the watermarks are pattern-based~\cite{wang2019neural,gao2019strip,chen2019deepinspect,guo2019tabor}. In this work, we study fine-tuning as a generic and effective approach to watermark removal, without the knowledge of the specific watermarking scheme.

\paragraph{Limited data for fine-tuning.} We assume that the adversary has computation resources for fine-tuning, and this assumption is also made in previous work studying fine-tuning and distillation-based approaches for watermark removal~\cite{adi2018turning,zhang2018protecting,liu2018fine,yang2019effectiveness}. Note that most prior works along this line assume that the adversary has access to the same amount of benign data for task $\mathcal{T}$ as the model owner. However, this assumption does not always hold in reality. Specifically, when the adversary has a sufficiently large dataset to train a good model, such an adversary is generally less motivated to take the risk of conducting watermark removal attacks, given that the adversary is already able to train his own model from scratch.

To study the watermark removal problem with a more realistic threat model, in this work, we perform a comprehensive study of the scenarios where the adversary has a much smaller dataset for fine-tuning than the pre-training dataset. In this case, training a model from scratch with such a limited dataset would typically result in inferior performance, as we will demonstrate in Section~\ref{sec:eval}, which provides the adversary with sufficient incentives to pirate a pre-trained model and invalidate its watermarks.

\section{REFIT: REmoving watermarks via FIne-Tuning}
\label{sec:method}

\begin{figure}[h]
    \centering
    \includegraphics[width=\linewidth]{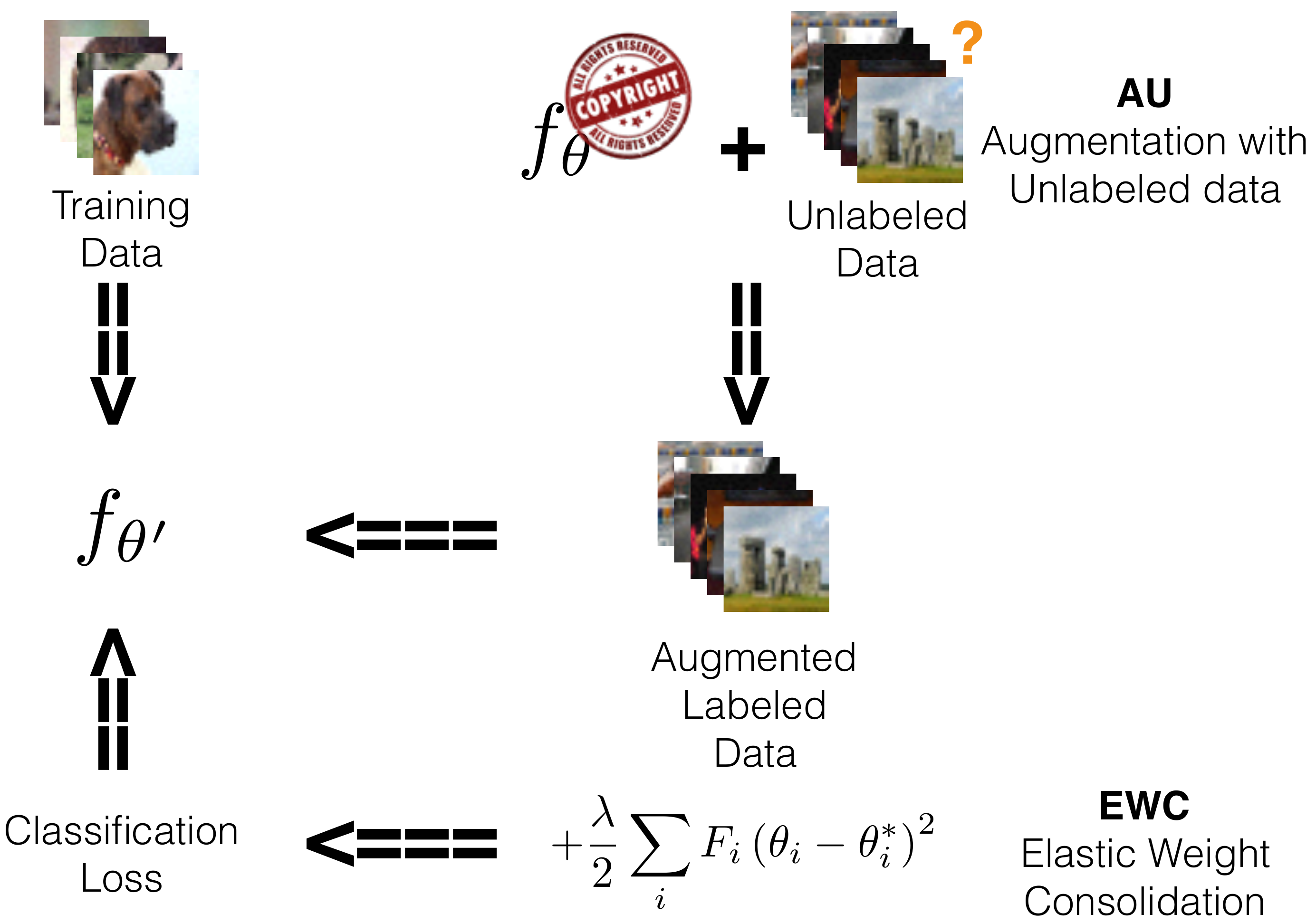}
    \caption{An overview of our proposed {\ours} framework. Specifically, besides the basic fine-tuning scheme, {\ours} incorporates two techniques to address the challenge when the adversary has a limited amount of in-distribution labeled data, i.e., elastic weight consolidation (EWC) and augmentation with unlabeled data (AU).}
    \label{fig:refit}
\end{figure}

In this section, we present {\ours}, a unified watermark removal framework based on fine-tuning. We present an overview of the framework in Figure~\ref{fig:refit}, and we will discuss the technical details in the later part of the section. The central intuition behind this scheme stems from the~\emph{catastrophic forgetting} phenomenon of machine learning models, that is, when a model is trained on a series of tasks, such a model could easily forget how to perform the previously trained tasks after training on a new task~\cite{goodfellow2013empirical,kirkpatrick2017overcoming,kemker2018measuring}. Accordingly, when the adversary further trains the model with his own data during the fine-tuning process, since the fine-tuning data no longer includes the watermark samples, the model should forget the previously learned watermark behavior.

Contrary to this intuition, some prior works show that existing watermarking techniques are robust against fine-tuning based techniques, even if the adversary fine-tunes the entire model and has access to the same benign data as the owner, i.e., the entire data for pre-training excluding the watermark samples~\cite{adi2018turning,zhang2018protecting,liu2018fine}. The key reason could be that the fine-tuning learning rates set in these works are too small to change the model weights with a small number of training epochs. To confirm this hypothesis, we first replicate the experiments in~\cite{adi2018turning} to embed watermarks into models trained on CIFAR-10 and CIFAR-100 respectively. Afterwards, we fine-tune the models in a similar way as their FTAL process, i.e., we update the weights of all layers. The only change is that instead of setting a small learning rate for fine-tuning, which is $0.001$ in their evaluation, we vary the magnitude of the learning rate to see its effect. Specifically, starting from 1e-5, the learning rate is doubled every 20 epochs in the fine-tuning process, which is the number of fine-tuning epochs for watermark removal in their evaluation.

\begin{figure}[t]
    \centering
    \includegraphics[width=\linewidth]{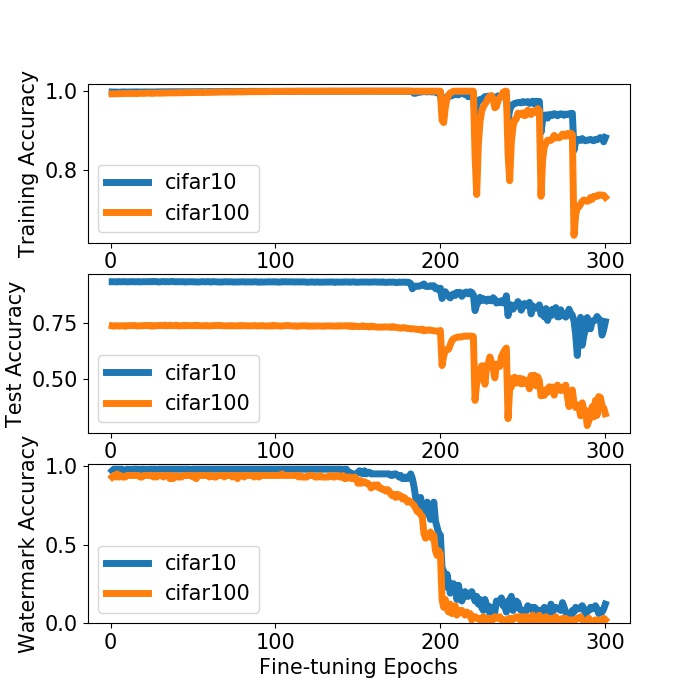}
    \caption{Training curves to illustrate the effect of learning rate during the fine-tuning stage, using $20\%$ of the labeled training data. At the beginning, the model is pre-trained with the watermark scheme in~\cite{adi2018turning}. Starting from a fine-tuning learning rate of 1e-5, the learning rate is doubled every 20 epochs. The watermark accuracy considerably decreases only when the learning rate is appropriately large. See Figure~\ref{fig:double-lr-20} in the appendix for the corresponding plots where the model is fine-tuned on the entire training set, by which we can draw similar conclusions.}
    \label{fig:double-lr-partial02-20}
\end{figure}

Figure~\ref{fig:double-lr-partial02-20} presents the training curve of this fine-tuning process. We can observe that the change of model performance is still negligible when the learning rate is around $0.001$, becomes noticeable when it reaches around $0.005$, and requires a larger value to reach a sufficiently low watermark accuracy. Meanwhile, at the beginning of each epoch when the learning rate is doubled, the training and test accuracies decrease first, then gradually improve within the next 20 epochs; on the other hand, the watermark accuracy keeps decreasing, since the watermarks are not included in the fine-tuning dataset. Therefore, although the adversary does not have knowledge of the watermarks, the adversary can set the initial fine-tuning learning rate so that it considerably degrades the training and test accuracies within the first few fine-tuning steps, which suggests that the model weights are sufficiently modified to remove the watermarks; on the other hand, desirable test performance is achieved when the fine-tuning converges, meaning that the initial learning rate does not have to be so large that results in a model not much different from one trained from scratch. In Section~\ref{sec:eval}, we demonstrate that with a learning rate schedule designed in this way, the adversary is able to remove the watermarks without compromising the model performance, when the adversary has access to a large amount of labeled training data.

While this initial attempt of watermark removal is promising, this basic fine-tuning scheme is inadequate when the adversary does not have training data comparable to the owner of the watermarked model. For example, when the adversary only has 20\% of the CIFAR-100 training set, to ensure that the watermarks are removed, the test accuracy of the fine-tuned model could degrade by $5\%$. This is again due to the catastrophic forgetting: when we fine-tune the model to forget its predictions on the watermark set, the model also forgets part of the normal training samples drawn from the same distribution as the test one. Although the decrease of the test accuracy is in general much less significant than watermark accuracy, such degradation is still considerable, which could hurt the utility of the model.

There have been some attempts to mitigate the catastrophic forgetting phenomenon in the literature~\cite{kirkpatrick2017overcoming,coop2013ensemble}. However, most techniques are not directly applicable to our setting. In fact, during the watermark embedding stage, the model is jointly trained on two tasks: (1) to achieve a good performance on a task of interest, e.g., image classification on CIFAR-10; (2) to remember the labels of images in the watermark set. Contrary to previous studies of catastrophic forgetting, which aims at preserving the model's predictions on all tasks it has been trained, our goal of watermark removal is two-fold, i.e., minimizing the model's memorization on the watermark task, while still preserving the performance on the main task it is evaluated on. This conflict results in the largest difference between our watermark removal task and the continual learning setting studied in previous work.

Another important difference is that although the training data of the adversary is different from the pre-trained data, the fine-tuning dataset contributes to a sub-task of the pre-trained model, while getting rid of the watermarks. On the other hand, different tasks are often complementary with each other in previous studies of catastrophic forgetting. This key observation enables us to adapt elastic weight consolidation~\cite{kirkpatrick2017overcoming}, a regularization technique proposed to mitigate the catastrophic forgetting issue, for our purpose of watermark removal.

\paragraph{Elastic Weight Consolidation (EWC).} The central motivation of EWC is to slow down the learning of parameters that are important for previously trained tasks~\cite{kirkpatrick2017overcoming}. To measure the contribution of each model parameter to a task, EWC first computes the diagonal of the Fisher information matrix of the previous task as follows:
\begin{align}
    F_i = F_{ii} = \mathbb{E}_{x \sim D, y \sim f_{\theta^{*}}(y|x)} \left[  
                \left.
                \frac{\partial \log f_{\theta}(y|x)}{\partial \theta_i} 
                \right|_{\theta = \theta^{*}}^2
        \right]
    \label{eq:EWC-Fisher}
\end{align}
where $f_{\theta^{*}}(y|x)$ is the probability distribution obtained by applying the softmax to output logits of the model with parameters $\theta^{*}$ given an input $x$, and $D$ is the training dataset of the previous task. The entire Fisher information matrix is given by $
F_{ij} = \mathbb{E}_{x \sim D, y \sim f_{\theta^{*}}(y|x)} \left[  
                \left.
                \frac{\partial \log f_{\theta}(y|x)}{\partial \theta_i} \cdot
                \frac{\partial \log f_{\theta}(y|x)}{\partial \theta_j} 
                \right|_{\theta = \theta^{*}}
        \right]
$, which defines a Riemannian metric on the parameter space.

Intuitively, to prevent the model from forgetting prior tasks when learning a new task, the learned parameter $\theta$ should be close to the parameter $\theta^{*}$ of prior tasks, when the new data also contains information relevant to $\theta^{*}$. Algorithmically, we penalize the distance between $\theta_i$ and $\theta^{*}_i$ when the $i$-th diagonal entry of the Fisher information matrix is large. Specifically, EWC adds a regularization term into the loss function for training on a new task, i.e.,
\begin{align}
    \mathcal{L}_{EWC}(\theta) = \mathcal{L}_{basic}(\theta) + \frac{\lambda}{2}\sum_{i} F_i \left( \theta_i - \theta_i^{*} \right)^2 
\label{eq:EWC-loss}
\end{align}
where $\mathcal{L}_{basic}(\theta)$ is the loss to optimize the performance on the new task (e.g. a cross entropy loss); $\lambda$ controls the strength of the regularization, indicating the importance of memorizing old tasks; $\theta^{*}$ is the parameters trained with the previous task; $F$ is the Fisher information matrix associated with $f_{\theta^{*}}$, and $F_i$ is the diagonal entry corresponding to the i-th parameter.

We can further extend this idea to the transfer learning setting, when the fine-tuning data belongs to a different task from the pre-trained one. In this case, the adversary can first fine-tune the pre-trained watermarked model with a small learning rate, which results in a model for his new task, although the watermarks usually still exist. Afterwards, the adversary can treat the model parameters of this new model as $\theta^{*}$, and plug in Equation~\ref{eq:EWC-Fisher} correspondingly.

Notice that since we do not have access to the pre-trained data, in principle we are not able to compute the Fisher information matrix of the previous task, thus cannot calculate the regularization term in $\mathcal{L}_{EWC}(\theta)$. However, by leveraging the assumption that the fine-tuning data is drawn from a similar distribution to the pre-trained data, we can instead approximate the Fisher matrix using the fine-tuning data. Given that the fine-tuning data contains no watermark, the EWC component enables the model to update less on model weights important for achieving a good test accuracy, while the model weights important for watermark memorization are still sufficiently modified. Although the approximation could be imprecise in this way, in Section~\ref{sec:eval}, we will show that this technique enables the adversary to improve the test performance of the model with limited data, while the watermarks are successfully removed.

With the same goal of preserving the test performance of the model with watermarks removed, we propose data augmentation with unlabeled data, which further decreases the amount of in-distribution labeled training samples needed for this purpose.

\paragraph{Augmentation with Unlabeled data (AU).} We propose to augment the fine-tuning data with unlabeled samples, which could easily be collected from the Internet. Let $\mathcal{U}=\{x_u\}_{u=1}^U$ be the unlabeled sample set, we use the pre-trained model as the labeling tool, i.e., $y_u=f_\theta(x_u)$ for each $x_u \in \mathcal{U}$. We have tried more advanced semi-supervised techniques to utilize the unlabeled data, e.g., virtual adversarial training~\cite{miyato2018virtual} and entropy minimization~\cite{grandvalet2005semi}, but none of them provides a significant gain compared to the aforementioned simple approach. Therefore, unless otherwise specified, we use this method for our evaluation of unlabeled data augmentation. Similar to our discussion of extending EWC to transfer learning, we can also apply this technique to the transfer learning setting by first fine-tuning the model for the new task without considering watermark removal, then using this model for labeling.

Since the test accuracy of the pre-trained model is not $100\%$ itself, such label annotation is inherently noisy; in particular, when $\mathcal{U}$ is drawn from a different distribution than the task of consideration, the assigned labels may not be meaningful at all. Nevertheless, they still enable the fine-tuned model to better mimic the prediction behavior of the pre-trained model. In Section~\ref{sec:eval}, we will show that leveraging unlabeled data significantly decreases the in-distribution labeled samples needed for effective watermark removal, while preserving the model performance.

\section{Evaluation Setup}
\label{sec:eval-setup}

In this section, we introduce the benchmarks and the watermark embedding schemes used in our evaluation, and discuss the details of our experimental configurations.

\subsection{Datasets}
\label{sec:datasets}

We evaluate on CIFAR-10~\cite{krizhevsky2009learning}, CIFAR-100~\cite{krizhevsky2009learning}, STL-10~\cite{coates2011analysis} and ImageNet32~\cite{chrabaszcz2017downsampled}, which are popular benchmarks for image classification, and some of them have been widely used in previous work on watermarking ~\cite{adi2018turning,zhang2018protecting,namba2019robust}.

\paragraph{CIFAR-10.} CIFAR-10 includes coloured images of 10 classes, where each of them has 5,000 images for training, and 1,000 images for testing. Each image is of size $32 \times 32$.

\paragraph{CIFAR-100.} CIFAR-100 includes coloured images of 100 classes, where each of them has 500 images for training, and 100 images for testing, thus the total number of training samples is the same as CIFAR-10. The size of each image is also $32 \times 32$.

\paragraph{STL-10.} STL-10 has been widely used for evaluating transfer learning, semi-supervised and unsupervised learning algorithms, featured with a large number of unlabeled samples. Specifically, STL-10 includes 10 labels, where each label has 500 training samples and 800 test samples. Besides the labeled samples, STL-10 also provides 100,000 unlabeled images drawn from a similar but broader distribution of images, i.e., they include images of labels that do not belong to the STL-10 label set. The size of each image is $96 \times 96$, which is much larger than CIFAR-10 and CIFAR-100. Although the label set of STL-10 and CIFAR-10 largely overlap, the images from them are distinguishable, even if resizing them to the same size.

\paragraph{ImageNet32.} ImageNet32 is a downsampled version of the ImageNet dataset~\cite{deng2009imagenet}. Specifically, ImageNet32 includes all samples in the training and validation sets of the original ImageNet, except that the images are resized to $32 \times 32$. Same as the original ImageNet, this dataset has 1.28 million training samples in 1000 classes, and 50,000 samples with 50 images per class for validation.

\subsection{Watermarking Techniques}

To demonstrate the effectiveness of {\ours} against various watermark embedding schemes, we evaluate pattern-based techniques~\cite{zhang2018protecting,chen2017targeted,gu2017badnets}, embedding samples drawn from other data sources as the watermarks~\cite{adi2018turning,zhang2018protecting,chen2017targeted}, exponential weighting~\cite{namba2019robust}, and adversarial frontier stitching~\cite{merrer2017adversarial}. These techniques represent the typical approaches of watermark embedding studied in the literature, and are shown to be the most effective ones against watermark removal.

\begin{figure}[t]
    \centering
    \begin{subfigure}[t]{0.15\linewidth}
    \includegraphics[width=\linewidth]{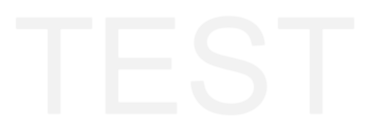}
    \caption{}
    \label{}
    \end{subfigure}
    \begin{subfigure}[t]{0.15\linewidth}
    \includegraphics[width=\linewidth]{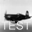}
    \caption{}
    \label{}
    \end{subfigure}
    \begin{subfigure}[t]{0.15\linewidth}    
    \includegraphics[width=\linewidth]{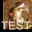}
    \caption{}
    \label{}
    \end{subfigure}
    \begin{subfigure}[t]{0.2\linewidth}
    \includegraphics[width=\linewidth]{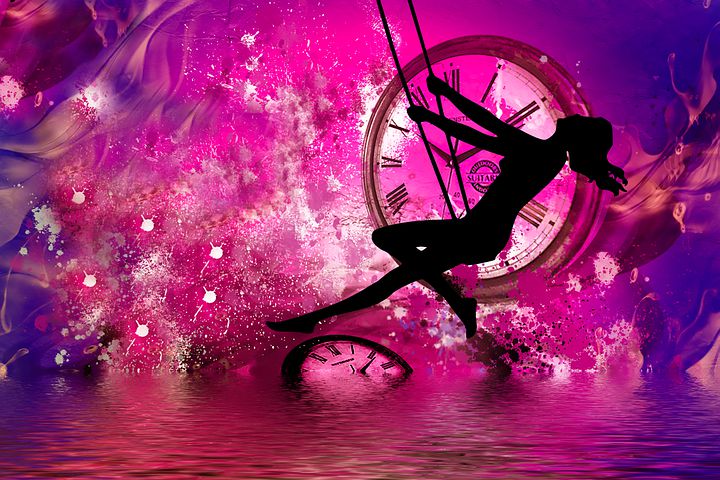}
    \caption{}
    \label{}
    \end{subfigure}
    \begin{subfigure}[t]{0.2\linewidth}
    \includegraphics[width=\linewidth]{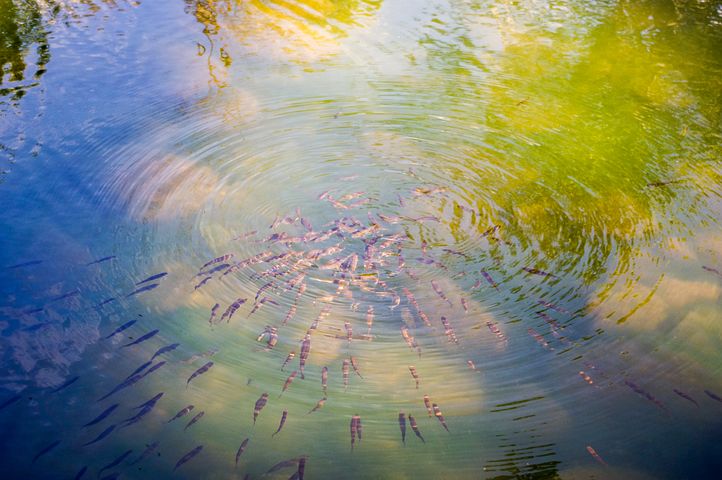}
    \caption{}
    \label{}
    \end{subfigure}
    \vspace{-0.5em}
    \caption{(a), (b), and (c) are examples of watermarks generated by the pattern-based technique in~\cite{zhang2018protecting}. Specifically, after an image is blended with the ``TEST'' pattern in (a), such an image is classified as the target label, e.g., an ``automobile'' on CIFAR-10. (d) and (e) are examples of watermarks generated by the out-of-distribution watermark embedding technique in~\cite{adi2018turning}, where different watermarks could have different assigned labels.}
    \label{fig:pattern-ood-watermark-ex}
    \vspace{-0.5em}
\end{figure}

\paragraph{Pattern-based techniques (Pattern).} A pattern-based technique specifies a key pattern $key$ and a target label $y^t$, so that for any image $x$ blended with the pattern $key$, $Pr(f_\theta(x)=y^t)$ is high. To achieve this, the owner generates a set of images $\{x^k\}_{k=1}^K$ blended with $key$, assigns $y^k=y^t (k \in 1, ..., K)$, then adds $\{(x^k, y^k)\}_{k=1}^K$ into the training set. See Figure~\ref{fig:pattern-ood-watermark-ex} for some watermark samples.

\paragraph{Out-of-distribution watermark embedding (OOD).} A line of work has studied using images from other data sources than the original training set as the watermarks. Figure~\ref{fig:pattern-ood-watermark-ex} presents some watermarks used in~\cite{adi2018turning}, where each watermark image is independently assigned with a random label, thus different watermarks can have different labels. Meanwhile, these images are very different from the samples in any benchmark we evaluate on, and do not belong to any category in the label set.

\paragraph{Exponential weighting (EW).} Compared to the above watermarking techniques, exponential weighting introduces two main different design choices~\cite{namba2019robust}. The first choice is about the watermark generation. Specifically, they change the labels of some training samples to different random labels, but do not modify the images themselves. The main motivation is to defend against the detection attacks mentioned in Section~\ref{sec:intro}, i.e., an adversary who steals the model could use an outlier detection scheme to detect input images that are far from the data distribution of interest, so as to bypass the watermark verification.

The second choice is about the embedding method. Instead of jointly training the model on both the normal training set and the watermark set, they decompose the training process into three stages. They first train the model on the normal training set only. Afterwards, they add an exponential weight operator over each model parameter. Specifically, for parameters in the $l$-th layer of the model denoted as $\theta^l$, $EW(\theta^l, T)_i = \frac{\exp{\left| \theta^l_i T \right| }}{\max_j \exp{\left| \theta^l_j T \right|}} \theta^l_i$, where $T$ is a hyper-parameter for adjusting the intensity of weighting. Finally, the model with the exponential weighting scheme is further trained on both normal training data and watermarks.

\begin{figure}[t]
    \centering
    \begin{subfigure}[t]{0.2\linewidth}
    \includegraphics[width=\linewidth]{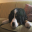}
    \caption{}
    \label{}
    \end{subfigure}
    \begin{subfigure}[t]{0.2\linewidth}
    \includegraphics[width=\linewidth]{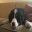}
    \caption{}
    \label{}
    \end{subfigure}
    \begin{subfigure}[t]{0.2\linewidth}
    \includegraphics[width=\linewidth]{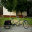}
    \caption{}
    \label{}
    \end{subfigure}
    \begin{subfigure}[t]{0.2\linewidth}
    \includegraphics[width=\linewidth]{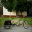}
    \caption{}
    \label{}
    \end{subfigure}
    \caption{Examples of watermarks generated by exponential weighting~\cite{namba2019robust} and adversarial frontier stitching~\cite{merrer2017adversarial}. (a) and (c) are generated by exponential weighting, which are images from the ImageNet32 training set, but assigned with random labels different from the ground truth; for example, the watermark label of (a) is ``trash can''. (b) and (d) are generated by adversarial frontier stitching technique, which add adversarial perturbations over (a) and (c) respectively, but keep the ground truth classes as their watermark labels; for example, the watermark label of (b) is still ``dog''.}
    \label{fig:ew-afs-watermark-ex}
    \vspace{-0.5em}
\end{figure}

\paragraph{Adversarial frontier stitching (ADV).} In~\cite{merrer2017adversarial}, they propose to use images added with the adversarial perturbation as the watermarks. Specifically, the model is first trained on the normal training set only. Afterwards, they generate a watermark set that is made up of 50\% true adversaries, i.e., adversarially perturbed images that the model provides the wrong predictions, and 50\% false adversaries, i.e., adversarially perturbed images on which the model still predicts the correct labels. The adversarial perturbations are computed using the fast gradient sign method~\cite{goodfellow2014explaining}, i.e., $x^{\text{adv}} = x + \epsilon \cdot sign(\nabla_x J(\theta, x, y))$, where $J(\theta, x, y)$ is the training loss function of the model, and $\epsilon$ controls the scale of the perturbation. Each of these images is annotated with the ground truth label of its unperturbed counterpart as its watermark label, i.e., the label of $x^{\text{adv}}$ is $y$, no matter whether it is a true adversary or false adversary. Finally, the model is fine-tuned with these watermarks added into the training set. See Figure~\ref{fig:ew-afs-watermark-ex} for examples of watermarks generated by this technique.

\vspace{-1em}
\subsection{Attack Scenarios}
\label{sec:attack-scenarios}

We consider the following attack scenarios in our evaluation.

\paragraph{Non-transfer learning.} The adversary leverages a watermarked model pre-trained for the same task as what the adversary desires. For this scenario, we evaluate on CIFAR-10, CIFAR-100, and ImageNet32. For CIFAR-10 and CIFAR-100, the watermarked model is pre-trained on its entire training set; for ImageNet32, the pre-trained model uses images of labels smaller than 500 in the training set. To simulate the scenario where the adversary can only collect a relatively small number of labeled training samples drawn from a similar distribution to the pre-training data, we vary the proportion of training samples the adversary has access to in the entire training set, but in practice, the fine-tuning training dataset does not necessarily need to be a subset of the pre-training dataset. We consider two data sources with abundant images for unlabeled data augmentation: (1) the unlabeled part of STL-10, which includes 100,000 samples; (2) for classification on CIFAR-10 and CIFAR-100, we also use the entire ImageNet32 for unlabeled data augmentation. For classification on ImageNet32, only those training samples with labels larger than 500 are included for unlabeled data augmentation. In both cases, we discard the labels of these ImageNet32 samples, and only use the images for augmentation. Note that these unlabeled images are very different from the labeled data. In particular, the label sets between CIFAR-100 and STL-10 barely overlap, and the label set of ImageNet32 is much more fine-grained than CIFAR-10 and CIFAR-100.

\paragraph{Transfer learning.} The adversary leverages a watermarked model pre-trained for a different task from what the adversary desires. For this scenario, our evaluation is centered on achieving good performance on STL-10. Note that the labeled part of STL-10 only includes 5,000 samples, which is insufficient for training a model with high accuracy. Therefore, an adversary can leverage the pre-trained model on another task with a larger training set, then fine-tune the model on STL-10. This fine-tuning method is widely adopted for transfer learning~\cite{yosinski2014transferable}, and is also evaluated in~\cite{adi2018turning}. In particular, we perform the transfer learning to adapt from a model trained on CIFAR-10 or ImageNet32 to STL-10. We do not consider CIFAR-100 in this setting, because we find that adapting from a pre-trained CIFAR-100 model results in inferior performance on STL-10 compared to CIFAR-10 and ImageNet32, e.g., the accuracy on STL-10 is around $5\%$ lower than the model pre-trained on CIFAR-10, as presented in~\cite{adi2018turning}. We perform the unlabeled data augmentation in the same way as the non-transfer learning setting.

\subsection{Implementation Details}
\label{sec:implementation-details}

\paragraph{Watermarking schemes.} Our watermarking scheme configuration largely follows the same setups as their original papers. We tune the hyper-parameters to ensure that the pre-trained model achieves 100\% watermark accuracy for each scheme, i.e., for all embedded watermark samples, the model prediction is the same as the assigned watermark label. Also, the prediction confidence scores for these watermark samples are high, e.g., above 0.85, suggesting the strength of the embedded watermark. We directly use their open-source implementation when applicable. Specifically:

\begin{itemize}[leftmargin=*,noitemsep,topsep=0em]
\item \emph{Pattern-based techniques.} We use the text pattern in~\cite{zhang2018protecting}, and Figure~\ref{fig:pattern-ood-watermark-ex} presents some examples of generated watermarks. 

\item \emph{OOD watermark techniques.} The watermark images are from the code repository of~\cite{adi2018turning}~\footnote{\url{https://github.com/adiyoss/WatermarkNN}.}. The watermark set contains 100 individual images with labels randomly drawn from the entire label set, and Figure~\ref{fig:pattern-ood-watermark-ex} shows some examples.

\item \emph{Exponential weighting.} We set $T=2.0$ as in \cite{namba2019robust}. For each dataset, we use the last 100 samples from the training set to form the watermark set, and ensure that these watermark samples are never included in the fine-tuning training set.

\item \emph{Adversarial frontier stitching.} We set $\epsilon$ so that the watermark accuracy of a model trained without watermarks is around 50\%. The values of $\epsilon$ are 0.15, 0.10 and 0.05 for CIFAR-10, CIFAR-100 and ImageNet32 respectively.
\end{itemize}

\paragraph{Watermark removal techniques.} We always fine-tune the entire model for {\ours}, because we find that fine-tuning the output layer only is insufficient for watermark removal, as demonstrated in~\cite{adi2018turning}; moreover, it will completely fail to remove watermarks in the transfer learning setting by design. We have tried both FTAL and RTAL processes in~\cite{adi2018turning}. Specifically, FTAL directly fine-tunes the entire model; when using RTAL, the output layer is randomly initialized before fine-tuning. For non-transfer learning, we apply the FTAL method, as RTAL does not provide additional performance gain; for transfer learning, we apply the RTAL method, since the label sets of the pre-trained and fine-tuning datasets are different. We observe that as long as the pre-trained model achieves a high test accuracy and fits the watermarks well, the model architecture does not have a critical influence on the effectiveness of watermark embedding and removal. Thus, unless otherwise specified, we mainly apply the ResNet-18 model~\cite{he2016deep} in our evaluation, which is able to achieve competitive performance on all benchmarks in our evaluation.

As discussed in Section~\ref{sec:method}, the failure of previous fine-tuning-based watermark removal approaches is mainly due to the improper learning rate schedule for fine-tuning. For example, the initial learning rate for fine-tuning is $0.001$ in~\cite{adi2018turning}, which is $100\times$ smaller than the initial learning rate for pre-training. In our evaluation, we set the initial fine-tuning learning rate to be much larger, e.g., $0.05$. We used SGD as the optimizer and set the batch size to be 100 for both pre-training and fine-tuning without unlabeled data, following the setup in~\cite{adi2018turning}. We fine-tune the model until the training accuracy does not further improve, which is typically within 20 epochs, as in~\cite{adi2018turning}. For unlabeled data augmentation, when there is no in-distribution labeled samples, each batch includes 100 unlabeled samples. When fine-tuning on CIFAR-10, CIFAR-100 and STL-10, we decay the learning rate by 0.9 every 500 steps. When fine-tuning on partial ImageNet32, the learning rate is multiplied by $0.9^t$ after training on the $\frac{t}{10}$-fraction of the entire training set. More discussion on implementation details is in Appendix~\ref{app:exp}. In Section~\ref{sec:eval}, we denote this basic version of {\ours} without EWC and AU as~\emph{Basic}.

For our EWC component, Fisher information is approximated by drawing $M$ samples from in-distribution labeled data available to the adversary. Unless otherwise specified, we set $M=10,000$ when the target domain is CIFAR-10, CIFAR-100 or STL-10, and $M=40,000$ when the target domain is ImageNet32. Notice that the samples are drawn with replacement, so $M$ can be larger than the number of training examples available, where the same example may be used multiple times. In practice, to improve the stability of the optimization, we first normalize the Fisher matrix $F_i$ so that its maximum entry is $1$, then clip the matrix by $\frac{1}{\lambda \cdot lr}$ before plugging it into Equation (\ref{eq:EWC-loss}), where $lr$ is the learning rate.

In addition, we also compare with two baseline methods denoted as~\emph{FS}, which train the entire model from scratch, so that the model is guaranteed to have a watermark accuracy no higher than the decision threshold, though the test accuracy is typically sub-optimal, especially when the training data is limited. The basic version simply trains on the dataset available for fine-tuning, without leveraging the pre-trained model. The second variant, denoted as~\emph{AU}, applies the pre-trained model as the labeling tool in the same way as the AU module in {\ours}, but the model is randomly initialized instead of initializing from the pre-trained model.

\paragraph{Evaluation metrics.} We consider the two metrics below.

\begin{itemize}[leftmargin=*,noitemsep,topsep=0em]
\item \emph{Watermark accuracy.} The adversary needs to make sure that the model accuracy on the watermark set is no more than the watermark decision threshold $\gamma$, i.e., the model predictions are the same as the assigned watermark labels for no more than $\gamma$ of the watermark inputs. In particular, we set $\gamma$ to be within the range of watermark accuracies of models trained without watermarks. Specifically, we trained 10 models from different random initialization without watermarks, evaluated their watermark accuracies, and set the threshold to ensure that there is no false positive, i.e., the watermark accuracies of these models are not accidentally higher than $\gamma$. For watermark schemes other than ADV, we set $\gamma$ to be $20\%$ for CIFAR-10, $10\%$ for CIFAR-100, and $3\%$ for ImageNet32. We set $\gamma=58\%$ for all benchmarks when using ADV, following~\cite{merrer2017adversarial}.
    
Notice that for the transfer learning setting, due to the difference of the label sets between the pre-trained and fine-tuning tasks, the embedded watermarks naturally do not apply to the new model. To measure the watermark accuracy, following~\cite{adi2018turning}, we replace the output layer of the fine-tuned model with the original output layer of the pre-trained model.

\item \emph{Test accuracy.} The adversary also aims to maximize the model accuracy on the normal test set. We evaluate the top-1 accuracy.
\end{itemize}

Regarding the presentation of evaluation results in the next section, unless otherwise specified, we only present the test accuracies of the models. The watermark accuracy of the pre-trained model embedded with any watermarking scheme in our evaluation is $100\%$, and the watermark accuracy of the model after watermark removal using {\ours} is always below the threshold $\gamma$.
\section{Evaluation}
\label{sec:eval}

In this section, we demonstrate the effectiveness of {\ours} to remove watermarks embedded by several different schemes, in both transfer and non-transfer learning scenarios.

\vspace{-1em}
\subsection{Evaluation of transfer learning}
\label{sec:eval-transfer}

\begin{table*}[tbp]
\begin{tabular}{cc}
    \begin{minipage}{.43\linewidth}
    \centering
    \scalebox{0.9}{\begin{tabular}{c|cc|ccc}
    \toprule
    & \multicolumn{2}{c|}{\textbf{FS}} & \multicolumn{3}{c}{\textbf{\ours}} \\
     & \textbf{Basic} & \textbf{AU} & \textbf{Basic} & \textbf{EWC} & \textbf{AU} \\
    \midrule
    Pattern & \multirow{4}{*}{$66.15$} & $75.28$/$74.01$ & $82.96$ & $83.76$ & $83.80$/{\large $\mathbf{84.36}$} \\
    OOD & & $74.69$/$74.59$ & $82.83$ & {\large $\mathbf{83.90}$}& $83.51$/$83.40$ \\
    EW & & $75.51$/$74.48$ & $84.03$ & {\large $\mathbf{84.66}$} & $84.43$/$84.07$ \\
    ADV & & $75.23$/$73.95$ & $83.66$ & {\large $\mathbf{84.39}$} & {\large $\mathbf{84.39}$}/$83.80$  \\
    \bottomrule
    \end{tabular}}
    \caption{Test accuracies (\%) of models on STL-10 after watermark removal in the transfer learning setting, where the models are pre-trained on CIFAR-10. The accuracies of fine-tuned models on STL-10 with no requirement for watermark removal are $82.06\%$, $82.89\%$, $84.03 \%$ and $83.66\%$ for Pattern, OOD, EW and ADV respectively. For AU, x/y stands for the results of augmenting with STL-10 and ImageNet32 respectively.}
    \label{tab:res-cifar10-stl}
    \end{minipage}
    &
    \begin{minipage}{.45\linewidth}
    \centering
    \scalebox{0.8}{\begin{tabular}{c|cc|cccc}
    \toprule
    & \multicolumn{2}{c|}{\textbf{FS}} & \multicolumn{4}{c}{\textbf{\ours}} \\
     & \textbf{Basic} & \textbf{AU}  & \textbf{Basic} & \textbf{EWC} & \textbf{AU} & \textbf{EWC+AU} \\
    \midrule
    Pattern & \multirow{4}{*}{$66.15$} & $74.76$/$71.50$ & $88.89$ & $91.14$ & $92.30/90.78$& {\large $\mathbf{93.31}$}/$92.99$  \\
    OOD & & $75.63$/$72.50$ & $90.39$ & $92.03$ & $92.74/91.96$ & {\large $\mathbf{92.94}$}/$92.45$   \\
    EW & & $75.56$/$72.36$ & $91.01$ & $91.68$ & $92.11/91.41$ & {\large $\mathbf{92.46}$}/$92.34$  \\
    ADV & & $75.19$/$72.71$ & $92.46$ & $92.63$ & $92.63/92.51$ & {\large $\mathbf{92.96}$}/$92.65$ \\
    \bottomrule
    \end{tabular}}
    \caption{Test accuracies (\%) of models on STL-10 after watermark removal in the transfer learning setting, where the models are pre-trained on ImageNet32.  The accuracies of fine-tuned models on STL-10 with no requirement for watermark removal are $92.95\%$, $92.39\%$, $92.16\%$, and $92.46\%$ for Pattern, OOD, EW and ADV respectively. For AU, x/y stands for the results of augmenting with STL-10 and ImageNet32 respectively.}
    \label{tab:res-imgnet-stl}
    \end{minipage}
\end{tabular}
\end{table*}

\paragraph{Pre-training on CIFAR-10.} We first present the results of transfer learning from CIFAR-10 to STL-10 in Table~\ref{tab:res-cifar10-stl}. We observe that with the basic version of {\ours}, where neither EWC nor AU is applied, removing watermarks already does not compromise the model performance on the test set. When equipped with either EWC or AU, the model fine-tuned with {\ours} even surpasses the performance of the watermarked model.

\paragraph{Pre-training on ImageNet32.} The results of transferring from ImageNet32 to STL-10 are in Table~\ref{tab:res-imgnet-stl}. We observe that using the pre-trained models on ImageNet32 yields around $10\%$ improvement of test accuracy compared to the ones pre-trained on CIFAR-10, although the label set of ImageNet32 is much more different from STL-10 than CIFAR-10. This could attribute to the diversity of samples in ImageNet32, which makes it a desirable data source for pre-training. Different from pre-training on CIFAR-10, the basic version of {\ours} no longer suffices to preserve the test accuracy. By leveraging the unlabeled part of STL-10, the model performance becomes comparable to the watermarked ones. When combining EWC and AU, the performance of fine-tuned models dominates among different variants of {\ours} and the watermarked models.

\paragraph{Discussion of different pre-training datasets.} Meanwhile, when we train the STL-10 from scratch and only use the pre-trained model as the labeling tool, the performance of models fine-tuned on unlabeled part of STL-10 is consistently better than models using ImageNet32 for unlabeled data augmentation. This is expected since the unlabeled part of STL-10 is closer to the test distribution than ImageNet32. Interestingly, we find that by integrating AU into {\ours}, the gap between utilizing STL-10 and ImageNet32 for unlabeled data augmentation is significantly shrunk, which indicates the effectiveness of our overall framework.

\subsection{Evaluation of non-transfer learning}
\label{sec:eval-nontransfer}

\begin{table*}[tbp]
\begin{tabular}{cc}
\begin{minipage}{.43\linewidth}
    \centering
    \scalebox{0.9}{\begin{tabular}{c|cc|ccc}
    \toprule
    \multirow{2}{*}{\textbf{Pct.}} & \multicolumn{2}{c|}{\textbf{FS}} & \multicolumn{3}{c}{\textbf{\ours}} \\
    & \textbf{Basic} & \textbf{AU} & \textbf{Basic} & \textbf{EWC} & \textbf{AU} \\
    \midrule
    \multicolumn{6}{c}{\textbf{Pattern}} \\
    \midrule
    $0\%$ & $-$ & $89.86/88.43$ & $-$ &$-$ & {\large $\mathbf{92.53}$}/$91.93$ \\
    $20\%$ & $87.40$ & $91.32/90.91$ & $92.12$ & {\large $\mathbf{92.90}$} & $92.80/92.78$ \\
    $30\%$ & $89.64$ & $92.13/91.49$ & $92.22$ &$93.02$ & {\large $\mathbf{93.15}$}/$92.88$ \\
    $40\%$ & $90.46$ & $92.46/92.15$ & $92.93$ & {\large $\mathbf{93.25}$} & $93.18/93.03$ \\
    $50\%$ & $91.45$ & $92.47/92.25$ & $93.08$ & {\large $\mathbf{93.25}$} & $93.18/93.13$ \\
    $80\%$ & $93.01$ & $92.82/92.67$ & $93.52$ &$93.67$ & {\large $\mathbf{94.11}$}/$93.43$\\
    \midrule
    \multicolumn{6}{c}{\textbf{OOD}} \\
    \midrule
    $0\%$ & $-$ & $90.13/88.01$ & $-$ &$-$ & {\large $\mathbf{90.48}$}/$87.52$ \\
    $20\%$ & $87.40$ & $91.15/90.87$ & $91.19$ &$91.85$ & {\large $\mathbf{92.41}$}/$92.08$ \\
    $30\%$ & $89.64$ & $91.67/91.58$ & $91.58$ &$92.58$ & {\large $\mathbf{93.01}$}/$92.61$ \\
    $40\%$ & $90.46$ & $92.11/91.92$ & $92.76$ &$93.20$ & {\large $\mathbf{93.21}$}/$92.58$ \\
    $50\%$ & $91.45$ & $92.48/92.29$ & $92.97$ & {\large $\mathbf{93.37}$} & $93.21/92.66$ \\
    $80\%$ & $93.01$ & $92.81/92.66$ & $93.93$ &$93.85$ & {\large $\mathbf{94.00}$}/$93.26$ \\
    \midrule
    \multicolumn{6}{c}{\textbf{EW}} \\
    \midrule
    $0\%$ & $-$ & $89.77/89.11$ & $-$ &$-$ & $93.05$/{\large $\mathbf{93.22}$} \\
    $20\%$ & $87.40$ & $91.58/90.99$ & $91.65$ &$92.46$ &$93.30$/{\large $\mathbf{93.34}$} \\
    $30\%$ & $89.64$ & $91.69/91.69$ & $92.30$ &$93.29$ & {\large $\mathbf{93.50}$}/$93.39$ \\
    $40\%$ & $90.46$ & $92.35/91.92$ & $92.83$ &$93.27$ & $93.34$/{\large $\mathbf{93.42}$} \\
    $50\%$ & $91.45$ & $92.44/92.31$ & $93.39$ &$93.39$ & {\large $\mathbf{93.51}$}/$93.36$ \\
    $80\%$ & $93.01$ & $92.97/93.03$ & $93.95$ & {\large $\mathbf{94.05}$} & $93.61/93.42$ \\
    \midrule
    \multicolumn{6}{c}{\textbf{ADV}} \\
    \midrule
    $0\%$ & $-$ & $90.05/79.47$ & $-$ &$-$ & {\large $\mathbf{91.60}$}/$85.68$ \\
    $20\%$ & $87.40$ & $91.52/89.07$ & $92.85$ &$92.95$ & {\large $\mathbf{93.09}$}/$92.72$ \\
    $30\%$ & $89.64$ & $92.09/90.02$ & $93.16$ & {\large $\mathbf{93.40}$} & $93.09/93.01$ \\
    $40\%$ & $90.46$ & $92.23/91.15$ & $93.21$ & {\large $\mathbf{93.37}$} & $93.20/93.09$ \\
    $50\%$ & $91.45$ & $92.58/91.83$ & $93.12$ & {\large $\mathbf{93.56}$} & $93.19/93.42$ \\
    $80\%$ & $93.01$ & $92.93/92.69$ & $93.69$ & {\large $\mathbf{93.80}$} & $93.65/93.76$ \\
    \bottomrule
    \end{tabular}}
    \caption{Results of non-transfer learning setting on CIFAR-10. The first column is the percentage of the CIFAR-10 training set used for fine-tuning, and the rest columns show the accuracy (\%) on the test set. The test accuracy of the pre-trained model is $93.23\%$ for Pattern, $93.63\%$ for OOD, $93.49\%$ for EW, and $93.31\%$ for ADV. For AU, x/y stands for the results of augmenting with STL-10 and ImageNet32 respectively.}
    \label{tab:res-cifar10}
    \end{minipage}
    &
    \begin{minipage}{.45\linewidth}
    \centering
    \scalebox{0.9}{\begin{tabular}{c|cc|ccc}
    \toprule
    \multirow{2}{*}{\textbf{Pct.}} & \multicolumn{2}{c|}{\textbf{FS}} & \multicolumn{3}{c}{\textbf{\ours}} \\
    & \textbf{Basic} & \textbf{AU} & \textbf{Basic} & \textbf{EWC} & \textbf{AU} \\
    \midrule
    \multicolumn{6}{c}{\textbf{Pattern}} \\
    \midrule
    $0\%$ & $-$ & $58.07/62.44$ & $-$ &$-$ & {\large $\mathbf{70.75}$}/$68.27$ \\
    $20\%$ & $56.72$ & $67.28/68.12$ & $68.88$ &$71.80$ & $71.97$/{\large $\mathbf{72.06}$} \\
    $30\%$ & $62.20$ & $68.95/70.07$ & $71.05$ &$72.64$ & {\large $\mathbf{72.98}$}/$72.73$ \\
    $40\%$ & $65.42$ & $70.45/71.34$ & $71.96$ &$73.20$ & {\large $\mathbf{73.44}$}/$73.39$ \\
    $50\%$ & $68.18$ & $71.27/72.23$ & $72.58$ &$73.44$ & $73.72$/{\large $\mathbf{73.84}$}\\
    $80\%$ & $71.71$ & $73.22/73.79$ & $74.23$ &$74.77$ & {\large $\mathbf{75.42}$}/$74.09$\\
    \midrule
    \multicolumn{6}{c}{\textbf{OOD}} \\
    \midrule
    $0\%$ & $-$ & $57.22/61.11$ & $-$ &$-$ & $65.98$/{\large $\mathbf{66.79}$} \\
    $20\%$ & $56.72$ & $67.18/67.75$ & $68.55$ &$69.91$ & {\large $\mathbf{71.02}$}/$71.00$ \\
    $30\%$ & $62.20$ & $68.83/70.06$ & $70.12$ &$71.77$ & $71.70$/ {\large $\mathbf{72.25}$} \\
    $40\%$ & $65.42$ & $70.44/71.10$ & $70.80$ & {\large $\mathbf{72.57}$} & $72.20/72.40$ \\
    $50\%$ & $68.18$ & $71.37/72.17$ & $72.27$ &$72.73$ & $72.73$/{\large $\mathbf{73.11}$} \\
    $80\%$ & $71.71$ & $72.65/73.00$ & $73.61$ & {\large $\mathbf{74.00}$} & $73.70/73.18$\\
    \midrule
    \multicolumn{6}{c}{\textbf{EW}} \\
    \midrule
    $0\%$ & $-$ & $55.79/64.35$ & $-$ &$-$ & $71.78$/ {\large $\mathbf{73.41}$} \\
    $20\%$ & $56.72$ & $67.66/68.57$ & $69.00$ &$70.63$ & {\large $\mathbf{73.48}$}/$73.34$ \\
    $30\%$ & $62.20$ & $69.01/70.71$ & $71.37$ &$72.13$ & $73.72$/ {\large $\mathbf{74.08}$} \\
    $40\%$ & $65.42$ & $70.72/71.30$ & $72.64$ &$73.27$ & $74.21$/ {\large $\mathbf{74.34}$} \\
    $50\%$ & $68.18$ & $71.96/72.38$ & $73.46$ &$74.25$ & $74.26$/ {\large $\mathbf{75.07}$} \\
    $80\%$ & $71.71$ & $73.70/73.56$ & $74.98$ & {\large $\mathbf{75.18}$} & $75.09/74.84$ \\
    \midrule
    \multicolumn{6}{c}{\textbf{ADV}} \\
    \midrule
    $0\%$ & $-$ & $57.47/64.89$ & $-$ &$-$ & {\large $\mathbf{69.92}$}/$68.64$ \\
    $20\%$ & $56.72$ & $67.22/67.81$ & $71.16$ &$71.46$ & {\large $\mathbf{71.67}$}/$71.58$ \\
    $30\%$ & $62.20$ & $69.30/69.40$ & $71.73$ &$72.20$ & {\large $\mathbf{72.28}$}/$72.02$ \\
    $40\%$ & $65.42$ & $70.74/71.31$ & $72.62$ & {\large $\mathbf{73.33}$} & $72.86/72.72$ \\
    $50\%$ & $68.18$ & $72.00/72.20$ & $73.01$ & {\large $\mathbf{73.41}$} & $73.11/73.26$ \\
    $80\%$ & $71.71$ & $72.71/73.01$ & $73.56$ & {\large $\mathbf{74.10}$} & $73.14/74.00$ \\
    \bottomrule
    \end{tabular}}
    \caption{Results of non-transfer learning setting on CIFAR-100. The first column is the percentage of the CIFAR-100 training set used for fine-tuning, and the rest columns show the accuracy (\%) on the test set. The test accuracy of the pre-trained model is $73.83\%$ for Pattern, $73.37\%$ for OOD, $74.95\%$ for EW, and $73.14\%$ for ADV. For AU, x/y stands for the results of augmenting with STL-10 and ImageNet32 respectively.}
    \label{tab:res-cifar100}
    \end{minipage}
\end{tabular}
\end{table*}

\paragraph{Results on CIFAR-10 and CIFAR-100.} For non-transfer learning setting, to begin with, we present results on CIFAR-10 and CIFAR-100 in Table~\ref{tab:res-cifar10} and~\ref{tab:res-cifar100} respectively. First, we observe that when the adversary has $80\%$ of the entire training set, using the basic version of {\ours} already achieves higher test accuracies than the pre-trained models using any watermarking scheme in our evaluation, while removing the watermarks. Note that the watermark accuracies are still above $95\%$ using the fine-tuning approaches in previous work~\cite{adi2018turning,zhang2018protecting}, suggesting the effectiveness of our modification of the fine-tuning learning rate schedule.

However, when the adversary only has a small proportion of the labeled training set, the test accuracy could degrade. Although the test accuracy typically drops for about $2\%$ on CIFAR-10 even if the adversary has only $20\%$ of the entire training set, the accuracy degradation could be up to $5\%$ on CIFAR-100. For all watermarking schemes other than ADV, incorporating EWC typically improves the test accuracy for nearly $1\%$ on CIFAR-10, and up to $3\%$ on CIFAR-100, which are significant considering the performance gap to the pre-trained models. The improvement for ADV is smaller yet still considerable, partially because the performance of the basic fine-tuning is already much better than other watermarking schemes, which suggests that ADV could be more vulnerable to watermark removal, at least when the labeled data is very limited. By leveraging the unlabeled data, the adversary is able to achieve the same level of test performance as the pre-trained models with only $20\% ~\sim 30\%$ of the entire training set. In particular, in Table~\ref{tab:res-cifar100-break}, we demonstrate that AU significantly improves the performance for labels with the lowest test accuracies. We skip the results of combining EWC and AU on CIFAR-10 and CIFAR-100, since they are generally very close to the results of AU. However, we will demonstrate that the combination of EWC and AU provides observable performance improvement on ImageNet32, which is a more challenging benchmark. We defer more discussion of sample efficiency to Appendix~\ref{app:eval-sample-efficiency}.

\paragraph{The effectiveness of AU.} Furthermore, unlabeled data augmentation enables the adversary to fine-tune the model without any labeled training data, and by solely relying on the unlabeled data, the accuracy of the fine-tuned model could be within $1\%$ difference from the pre-trained model on both CIFAR-10 and CIFAR-100, and sometimes even surpasses the performance of the model trained with $80\%$ data from scratch. Note that both STL-10 and ImageNet32 images are drawn from very different distributions than CIFAR-10 and CIFAR-100; when we only apply AU alone and train the model from scratch, the model accuracies are even worse than the basic version of {\ours}. Specifically, augmenting with STL-10 provides better results on CIFAR-10, partially because the label set of CIFAR-10 overlaps much more with STL-10 than ImageNet32; meanwhile, augmenting with ImageNet32 clearly shows better performance on CIFAR-100, which may result from its higher diversity that is necessary for CIFAR-100 classification. However, when integrating AU into {\ours}, the choice of unlabeled data does not play an important role in the final performance; i.e., the performance of augmenting with one data source is not always better than the other. These results show that {\ours} is effective without the requirement that the unlabeled data comes from the same distribution as the task of evaluation, which makes it a practical watermark removal technique for the adversary given its simplicity and efficacy, posing real threats to the robustness of watermark embedding schemes.

\paragraph{The effectiveness of EWC.} In addition, we notice that while AU mostly dominates when the percentage of labeled data is very small, with a moderate percentage of labeled data for fine-tuning, e.g., around $40\%$, EWC starts to outperform AU in some cases. In particular, on CIFAR-10, EWC typically becomes competitive to AU when $30\%$ labeled data is available to the adversary, and the corresponding percentage is $40\%$ on CIFAR-100. This indicates that with the increase of the labeled data, the estimated Fisher matrix could better capture the important model parameters to preserve for the adversary's task of interest.

\begin{table}[h]
\centering
    \begin{tabular}{|c|c|c|c|c|c|}
\hline
Basic&$ 34.00 $ &$ 37.00 $ &$ 39.00 $ &$ 42.00 $ &$ 43.00 $ \\
\hline
EWC&$ 32.00 $ &$ 45.00 $ &$ 40.00 $ &$ 49.00 $ &$ 44.00 $ \\
AU&$ {\large \mathbf{52.00}} $ &$ {\large \mathbf{46.00}} $ & {\large $\mathbf{46.00}$} & {\large $\mathbf{56.00}$} & {\large $\mathbf{54.00}$} \\
\hline
    \end{tabular}
    \caption{Results of non-transfer learning on CIFAR-100. We show the test accuracies (\%) for 5 labels with the lowest test accuracies. The pre-trained model is embedded with EW watermarks. Models are fine-tuned with $20\%$ of the CIFAR-100 training set. The AU uses STL-10 for data augmentation.}
    \label{tab:res-cifar100-break}
    \vspace{-1em}
\end{table}

\begin{table}[h]
    \centering
    \begin{tabular}{c|cc|cc|cc}
    \toprule
    \multirow{2}{*}{\textbf{Pct.}} & \multicolumn{2}{c|}{\textbf{FS}} & \multicolumn{3}{c}{\textbf{\ours}} \\
    & \textbf{Basic} & \textbf{AU} & \textbf{Basic} & \textbf{EWC} & \textbf{AU} & \textbf{EWC + AU}  \\
    \midrule
    \multicolumn{7}{c}{\textbf{Pattern}} \\
    \midrule
    $0\%$ & $-$ & $21.77$ & $-$ & $-$& \multicolumn{2}{c}{ {\large $\mathbf{54.37}$}} \\
    $10\%$ & $36.06$ & $39.41$ & $51.05$ & $53.59$ & $55.98$ & {\large $\mathbf{56.81}$} \\
    $20\%$ & $42.53$ & $48.34$ & $54.76$ & $56.35$ & $58.06$ & {\large $\mathbf{58.75}$} \\
    $30\%$ & $47.83$ & $52.76$ & $56.87$ & $58.40$ & $58.62$ & {\large $\mathbf{59.40}$} \\
    $40\%$ & $51.70$ & $55.24$ & $57.82$ & $59.09$ & $59.24$ & {\large $\mathbf{59.71}$} \\
    $50\%$ & $53.58$ & $57.04$ & $58.76$ & $59.68$ & $59.40$ & {\large $\mathbf{60.02}$} \\
    \midrule
    \multicolumn{7}{c}{\textbf{OOD}} \\
    \midrule
    $0\%$ & $-$ & $21.46$ & $-$ & $-$ &\multicolumn{2}{c}{{\large $\mathbf{51.68}$}} \\
    $10\%$ & $36.06$ & $39.32$ & $50.76$ & $52.02$ & $53.87$ & {\large $\mathbf{55.16}$} \\
    $20\%$ & $42.53$ & $48.30$ & $53.05$ & $54.64$ & $55.92$ & {\large $\mathbf{57.04}$} \\
    $30\%$ & $47.83$ & $52.58$ & $55.47$ & $56.42$ & $57.63$ & {\large $\mathbf{58.27}$} \\
    $40\%$ & $51.70$ & $55.34$ & $56.60$ & $57.41$ & $58.17$ & {\large $\mathbf{58.44}$} \\
    $50\%$ & $53.58$ & $56.87$ & $57.86$ & $58.50$ & $58.51$ & {\large $\mathbf{59.12}$} \\
    \midrule
    \multicolumn{7}{c}{\textbf{EW}} \\
    \midrule
    $0\%$ & $-$ & $23.56$ & $-$ & $-$ &\multicolumn{2}{c}{ {\large $\mathbf{52.76}$}} \\
    $10\%$ & $36.06$ & $39.70$ & $49.69$ & $52.44$ & $54.58$ & {\large $\mathbf{55.68}$} \\
    $20\%$ & $42.53$ & $48.16$ & $53.65$ & $55.89$ & $56.10$ & {\large $\mathbf{56.94}$} \\
    $30\%$ & $47.83$ & $52.26$ & $55.54$ & $56.25$ & $57.12$ & {\large $\mathbf{57.23}$} \\
    $40\%$ & $51.70$ & $55.32$ & $56.36$ & $57.00$ & $57.28$ & {\large $\mathbf{57.40}$} \\
    $50\%$ & $53.58$ & $56.90$ & $57.30$ & $57.68$ & $57.66$ & {\large $\mathbf{57.80}$} \\
    \midrule
    \multicolumn{7}{c}{\textbf{ADV}} \\
    \midrule
    $0\%$ & $-$ & $20.12$ & $-$ & $-$ & \multicolumn{2}{c}{ {\large $\mathbf{50.22}$}} \\
    $10\%$ & $36.06$ & $39.22$ & $50.27$ & $51.05$ & $53.52$ & {\large $\mathbf{53.72}$} \\
    $20\%$ & $42.53$ & $48.20$ & $52.95$ & $54.03$ & $56.00$ & {\large $\mathbf{56.50}$} \\
    $30\%$ & $47.83$ & $52.64$ & $55.21$ & $56.31$ & $57.02$ & {\large $\mathbf{57.40}$} \\
    $40\%$ & $51.70$ & $55.28$ & $57.43$ & $57.57$ & $57.90$ & {\large $\mathbf{57.94}$} \\
    $50\%$ & $53.58$ & $57.28$ & $57.88$ & $58.52$ & $58.02$ & {\large $\mathbf{58.83}$} \\
 \bottomrule
    \end{tabular}
    \caption{Results of non-transfer learning setting on ImageNet32. The first column is the percentage of the training set used for fine-tuning, and the rest columns show the test accuracy (\%). Note that the percentage is with respect to the training samples of the first 500 classes in ImageNet32. The test accuracy of the pre-trained model is $60.26\%$ for Pattern, $60.04\%$ for OOD, $58.31\%$ for EM, and $59.60\%$ for ADV. The reported test accuracy is measured on only the first 500 classes of ImageNet32. For AU, the unlabeled images are obtained from the last 500 classes of ImageNet32.}
    \label{tab:res-imgnet32}
    \vspace{-3em}
\end{table}

\paragraph{Results on ImageNet32.} In Table~\ref{tab:res-imgnet32}, we further present our results on ImageNet32. Compared to the results on CIFAR-10 and CIFAR-100, removing watermarks embedded into pre-trained ImageNet32 models could result in a larger decrease of test accuracy, which is expected given that ImageNet32 is a more challenging benchmark with a much larger label set. Despite facing more challenges, we demonstrate that by combining EWC and AU, {\ours} is still able to reach the same level of performance as the pre-trained watermarked model with 50\% of the labeled training data.

Meanwhile, the increased difficulty of this benchmark enables us to better analyze the importance of each component in {\ours}, i.e., EWC and AU. In particular, each of the two components offers a decent improvement of the test performance. The increase of accuracy with EWC is around $1\%-3\%$ over the basic version when the fine-tuning data is very limited, e.g., the percentage of labeled samples is $20\%$. The performance of using AU is generally better than using EWC, until the labeled training set includes $50\%$ of the ImageNet32 training samples of the first 500 classes, when EWC becomes more competitive. Finally, including both EWC and AU always enables further improvement of the test performance, suggesting that the combined technique is advantageous for challenging tasks.

\paragraph{Discussion of different watermarking schemes.} Comparing the results of different watermarking schemes, we observe that the models fine-tuned from pre-trained models embedded with pattern-based watermarks consistently beat the test accuracy of models embedded with other watermarks, suggesting that while pattern-based watermarking techniques are generally more often used than other approaches, especially for backdoor injection, such watermarks could be easier to remove, which makes it necessary to propose more advanced backdoor injection techniques that are robust to removal attacks. 

\subsection{Comparison with alternative watermark removal techniques}
\label{sec:eval-ab}

In the following, we provide some discussion and comparison with some general-purpose watermark removal approaches proposed in previous work, which also does not assume the knowledge of the concrete watermarking scheme.

\paragraph{Discussion of distillation-based approaches.} Distillation is a process to transform the knowledge extracted from a pre-trained model into a smaller model, while preserving the prediction accuracy of the smaller model so that it is comparable to the pre-trained one~\cite{hinton2015distilling}. Specifically, a probability vector is computed as $p(x)_i=\frac{exp(f(x)_i/T)}{\sum_{j}exp(f(x)_j/T)}$, where $f(x)$ is the output logit of the model $f$ given the input $x$, and $T$ is a hyper-parameter representing the temperature. Afterwards, instead of using the one-hot vector of the ground truth label for each training sample $x$, the extracted $p(x)$ from the pre-trained model is fed into the smaller model as the ground truth. Previous work has proposed distillation as a defense against adversarial examples~\cite{papernot2016distillation} and watermark embedding approaches~\cite{yang2019effectiveness}. Therefore, we investigate incorporating this technique into our framework.

Specifically, instead of using the one-hot encoding of labels predicted by the pre-trained model, we use $p(x)$ as the ground truth label and vary the value of $T$ to see the effect. Nevertheless, this method does not provide better performance; for example, with $20\%$ labeled training set on CIFAR-10 and using unlabeled part of STL-10 for augmentation, when the pre-trained model is embedded with OOD watermarks, setting $T=1$ provides a test accuracy of $91.60\%$, while using the one-hot label results in $91.93\%$ test accuracy as in Table~\ref{tab:res-cifar10}, and setting other values of $T$ do not cause any significant difference. In particular, we observe that when using output logits of the watermarked model as the ground truth for fine-tuning, the resulted model tends to have a higher watermark accuracy, perhaps because while the output logits allow the fine-tuned model to better fit the pre-trained model, it also encourages the fine-tuned model to learn more information of watermarks. Thus, we stick to our original design to annotate the unlabeled data.

\paragraph{Comparison with pruning and fine-pruning.} Previous work has studied the effectiveness of pruning-based approaches for watermark removal and found that such techniques are largely ineffective~\cite{zhang2018protecting,liu2018fine,namba2019robust}. In our evaluation, we observe that when applying the pruning approach, the watermark accuracy is tightly associated with the test accuracy, which makes it hard to find a sweet spot of the pruning rate so that the test performance is preserved while the watermarks are removed. We defer more discussion to Appendix~\ref{app:pruning}. On the other hand, prior work shows that combining pruning with fine-tuning could improve the effectiveness of watermark removal~\cite{liu2018fine}. Therefore, we also compare with the fine-pruning method proposed in~\cite{liu2018fine}, which first prunes part of the neurons that are activated the least for benign samples, and then performs the fine-tuning. We observe that {\ours} achieves a better test performance than the fine-pruning approach, even if we use the same learning rate schedule for both of them, which improves the performance of the fine-pruning algorithm compared to its original design. We defer more discussion of fine-pruning to Appendix~\ref{app:fine-pruning}.
\vspace{-0.5em}
\section{Related Work}

Aside from the attacks that infringe the intellectual property of a machine learning model, a variety of attacks have been proposed against machine learning models, aiming at either manipulating model predictions~\cite{gu2017badnets,shafahi2018poison,szegedy2013intriguing} or revealing sensitive information from trained models~\cite{shokri2017membership,hayes2019logan,fredrikson2014privacy,fredrikson2015model}. We will also review work on the catastrophic forgetting phenomenon in deep learning, as it inspires the use of EWC loss for our watermark removal scheme.

\paragraph{Model watermarking.} To protect the intellectual property of deep neural networks, prior works have proposed several watermarking schemes~\cite{zhang2018protecting,adi2018turning,namba2019robust,merrer2017adversarial}. For pattern-based techniques, watermark images are blended with the same pattern~\cite{zhang2018protecting}. Such techniques are also commonly used for backdoor attacks~\cite{chen2017targeted,gu2017badnets,liu2017trojaning}. Other watermark schemes utilize individual images as watermarks~\cite{adi2018turning,namba2019robust,merrer2017adversarial}. Existing watermark removal approaches are largely designed for pattern-based techniques~\cite{wang2019neural,gao2019strip,chen2019deepinspect,guo2019tabor}. Meanwhile, prior general-purpose watermark removal techniques require a large amount of training data to effectively remove the watermarks~\cite{yang2019effectiveness,liu2018fine}.

\paragraph{Backdoor attacks.} In the context of machine learning, backdoor attacks manipulate the model to provide the predictions specified by the adversary on inputs associated with the backdoor key. In this sense, backdoor attacks are closely connected to watermarks in their formats, but usually with different purposes, as discussed in~\cite{adi2018turning}. Previous work has shown that deep neural networks are vulnerable to backdoor attacks~\cite{chen2017targeted,gu2017badnets}. Accordingly, several defense methods have been proposed for backdoor attacks~\cite{wang2019neural,gao2019strip,chen2019deepinspect,guo2019tabor}.

\paragraph{Poisoning and evasion attacks.} Poisoning attacks inject well-crafted data into the training set to alter the predictive performance of a deep neural network. Besides backdoor injection, other attack goals include degrading the test accuracy indiscriminately, and changing the predictions of specific examples~\cite{biggio2012poisoning,nelson2008exploiting,li2016data,munoz2017towards,koh2017understanding,shafahi2018poison}. In contrast to poisoning attacks, evasion attacks are launched in the test time of a machine learning model. The resulted samples are called adversarial examples, which are visually similar to normal data but lead to wrong predictions by the model~\cite{biggio2013evasion,szegedy2013intriguing,goodfellow2014explaining,carlini2017towards}. Note that we leverage adversarial examples as the watermark for the ADV watermarking scheme.

\paragraph{Catastrophic forgetting.} Catastrophic forgetting refers to the phenomenon that a neural network model tends to underperform on old tasks when it is trained sequentially on multiple tasks. This occurs because the model weights that are
important for an old task are changed to meet the objectives of a new task. Many recent approaches have been proposed against this effect, such as adjusting weights~\cite{kirkpatrick2017overcoming,zenke2017continual}, and adding data of past tasks to the new task training~\cite{lopez2017gradient,shin2017continual}. In particular, elastic weight consolidation is a classic algorithm for mitigating catastrophic forgetting via adapting the learning of specific weights to their importance to previous tasks~\cite{kirkpatrick2017overcoming}. Note that the original EWC algorithm requires access to the data used for learning old tasks, which is not available in our case. Therefore, we propose an adaption of the algorithm to make it suitable for our watermark removal application.
\vspace{-0.5em}
\section{Conclusion}
\label{sec:conc}

In this work, we propose {\ours}, a unified framework that removes the watermarks via fine-tuning. We first demonstrate that by appropriately designing the learning rate schedule, our fine-tuning approach could effectively remove the watermarks. We further propose two techniques integrated into the {\ours} framework, i.e., an adaption of the elastic weight consolidation (EWC) approach, and unlabeled data augmentation (AU). We conduct an extensive evaluation with the assumption of a weak adversary who only has access to a limited amount of training data. Our results demonstrate the effectiveness of {\ours} against several watermarking schemes of different types. In particular, EWC and AU enable the adversary to successfully remove the watermarks without causing much degradation of the model performance.  Our study highlights the vulnerability of existing watermarking techniques, and we consider proposing more robust watermarking techniques as future work.

\ifarxiv
\section*{Acknowledgement}
This material is in part based upon work supported by the National 
Science Foundation under Grant No. TWC-1409915, Berkeley DeepDrive, and DARPA D3M under Grant No. FA8750-17-2-0091.
Any opinions, findings, and conclusions or recommendations expressed
in this material are those of the author(s) and do not necessarily
reflect the views of the National Science Foundation. Xinyun Chen is supported by the Facebook Fellowship.
\else
\fi

\bibliographystyle{ACM-Reference-Format}
\bibliography{ref}


\begin{thebibliography}{52}


\ifx \showCODEN    \undefined \def \showCODEN     #1{\unskip}     \fi
\ifx \showDOI      \undefined \def \showDOI       #1{#1}\fi
\ifx \showISBNx    \undefined \def \showISBNx     #1{\unskip}     \fi
\ifx \showISBNxiii \undefined \def \showISBNxiii  #1{\unskip}     \fi
\ifx \showISSN     \undefined \def \showISSN      #1{\unskip}     \fi
\ifx \showLCCN     \undefined \def \showLCCN      #1{\unskip}     \fi
\ifx \shownote     \undefined \def \shownote      #1{#1}          \fi
\ifx \showarticletitle \undefined \def \showarticletitle #1{#1}   \fi
\ifx \showURL      \undefined \def \showURL       {\relax}        \fi
\providecommand\bibfield[2]{#2}
\providecommand\bibinfo[2]{#2}
\providecommand\natexlab[1]{#1}
\providecommand\showeprint[2][]{arXiv:#2}

\bibitem[\protect\citeauthoryear{Adi, Baum, Cisse, Pinkas, and Keshet}{Adi
  et~al\mbox{.}}{2018}]%
        {adi2018turning}
\bibfield{author}{\bibinfo{person}{Yossi Adi}, \bibinfo{person}{Carsten Baum},
  \bibinfo{person}{Moustapha Cisse}, \bibinfo{person}{Benny Pinkas}, {and}
  \bibinfo{person}{Joseph Keshet}.} \bibinfo{year}{2018}\natexlab{}.
\newblock \showarticletitle{Turning your weakness into a strength: Watermarking
  deep neural networks by backdooring}. In \bibinfo{booktitle}{\emph{27th
  $\{$USENIX$\}$ Security Symposium}}.
\newblock


\bibitem[\protect\citeauthoryear{Biggio, Corona, Maiorca, Nelson,
  {\v{S}}rndi{\'c}, Laskov, Giacinto, and Roli}{Biggio et~al\mbox{.}}{2013}]%
        {biggio2013evasion}
\bibfield{author}{\bibinfo{person}{Battista Biggio}, \bibinfo{person}{Igino
  Corona}, \bibinfo{person}{Davide Maiorca}, \bibinfo{person}{Blaine Nelson},
  \bibinfo{person}{Nedim {\v{S}}rndi{\'c}}, \bibinfo{person}{Pavel Laskov},
  \bibinfo{person}{Giorgio Giacinto}, {and} \bibinfo{person}{Fabio Roli}.}
  \bibinfo{year}{2013}\natexlab{}.
\newblock \showarticletitle{Evasion attacks against machine learning at test
  time}. In \bibinfo{booktitle}{\emph{Joint European conference on machine
  learning and knowledge discovery in databases}}.
\newblock


\bibitem[\protect\citeauthoryear{Biggio, Nelson, and Laskov}{Biggio
  et~al\mbox{.}}{2012}]%
        {biggio2012poisoning}
\bibfield{author}{\bibinfo{person}{Battista Biggio}, \bibinfo{person}{Blaine
  Nelson}, {and} \bibinfo{person}{Pavel Laskov}.}
  \bibinfo{year}{2012}\natexlab{}.
\newblock \showarticletitle{Poisoning attacks against support vector machines}.
\newblock \bibinfo{journal}{\emph{arXiv preprint arXiv:1206.6389}}
  (\bibinfo{year}{2012}).
\newblock


\bibitem[\protect\citeauthoryear{Carlini and Wagner}{Carlini and
  Wagner}{2017}]%
        {carlini2017towards}
\bibfield{author}{\bibinfo{person}{Nicholas Carlini} {and}
  \bibinfo{person}{David Wagner}.} \bibinfo{year}{2017}\natexlab{}.
\newblock \showarticletitle{Towards evaluating the robustness of neural
  networks}. In \bibinfo{booktitle}{\emph{2017 IEEE Symposium on Security and
  Privacy (SP)}}.
\newblock


\bibitem[\protect\citeauthoryear{Chen, Carvalho, Baracaldo, Ludwig, Edwards,
  Lee, Molloy, and Srivastava}{Chen et~al\mbox{.}}{2018}]%
        {chen2018detecting}
\bibfield{author}{\bibinfo{person}{Bryant Chen}, \bibinfo{person}{Wilka
  Carvalho}, \bibinfo{person}{Nathalie Baracaldo}, \bibinfo{person}{Heiko
  Ludwig}, \bibinfo{person}{Benjamin Edwards}, \bibinfo{person}{Taesung Lee},
  \bibinfo{person}{Ian Molloy}, {and} \bibinfo{person}{Biplav Srivastava}.}
  \bibinfo{year}{2018}\natexlab{}.
\newblock \showarticletitle{Detecting backdoor attacks on deep neural networks
  by activation clustering}.
\newblock \bibinfo{journal}{\emph{arXiv preprint arXiv:1811.03728}}
  (\bibinfo{year}{2018}).
\newblock


\bibitem[\protect\citeauthoryear{Chen, Fu, Zhao, and Koushanfar}{Chen
  et~al\mbox{.}}{2019}]%
        {chen2019deepinspect}
\bibfield{author}{\bibinfo{person}{Huili Chen}, \bibinfo{person}{Cheng Fu},
  \bibinfo{person}{Jishen Zhao}, {and} \bibinfo{person}{Farinaz Koushanfar}.}
  \bibinfo{year}{2019}\natexlab{}.
\newblock \showarticletitle{DeepInspect: A Black-box Trojan Detection and
  Mitigation Framework for Deep Neural Networks}.
\newblock \bibinfo{journal}{\emph{International Joint Conferences on Artificial
  Intelligence (IJCAI)}} (\bibinfo{year}{2019}).
\newblock


\bibitem[\protect\citeauthoryear{Chen, Liu, Li, Lu, and Song}{Chen
  et~al\mbox{.}}{2017}]%
        {chen2017targeted}
\bibfield{author}{\bibinfo{person}{Xinyun Chen}, \bibinfo{person}{Chang Liu},
  \bibinfo{person}{Bo Li}, \bibinfo{person}{Kimberly Lu}, {and}
  \bibinfo{person}{Dawn Song}.} \bibinfo{year}{2017}\natexlab{}.
\newblock \showarticletitle{Targeted backdoor attacks on deep learning systems
  using data poisoning}.
\newblock \bibinfo{journal}{\emph{arXiv preprint arXiv:1712.05526}}
  (\bibinfo{year}{2017}).
\newblock


\bibitem[\protect\citeauthoryear{Chrabaszcz, Loshchilov, and Hutter}{Chrabaszcz
  et~al\mbox{.}}{2017}]%
        {chrabaszcz2017downsampled}
\bibfield{author}{\bibinfo{person}{Patryk Chrabaszcz}, \bibinfo{person}{Ilya
  Loshchilov}, {and} \bibinfo{person}{Frank Hutter}.}
  \bibinfo{year}{2017}\natexlab{}.
\newblock \showarticletitle{A downsampled variant of imagenet as an alternative
  to the cifar datasets}.
\newblock \bibinfo{journal}{\emph{arXiv preprint arXiv:1707.08819}}
  (\bibinfo{year}{2017}).
\newblock


\bibitem[\protect\citeauthoryear{Coates, Ng, and Lee}{Coates
  et~al\mbox{.}}{2011}]%
        {coates2011analysis}
\bibfield{author}{\bibinfo{person}{Adam Coates}, \bibinfo{person}{Andrew Ng},
  {and} \bibinfo{person}{Honglak Lee}.} \bibinfo{year}{2011}\natexlab{}.
\newblock \showarticletitle{An analysis of single-layer networks in
  unsupervised feature learning}. In \bibinfo{booktitle}{\emph{The fourteenth
  international conference on artificial intelligence and statistics}}.
\newblock


\bibitem[\protect\citeauthoryear{Coop, Mishtal, and Arel}{Coop
  et~al\mbox{.}}{2013}]%
        {coop2013ensemble}
\bibfield{author}{\bibinfo{person}{Robert Coop}, \bibinfo{person}{Aaron
  Mishtal}, {and} \bibinfo{person}{Itamar Arel}.}
  \bibinfo{year}{2013}\natexlab{}.
\newblock \showarticletitle{Ensemble learning in fixed expansion layer networks
  for mitigating catastrophic forgetting}.
\newblock \bibinfo{journal}{\emph{IEEE transactions on neural networks and
  learning systems}} (\bibinfo{year}{2013}).
\newblock


\bibitem[\protect\citeauthoryear{Deng, Dong, Socher, Li, Li, and Fei-Fei}{Deng
  et~al\mbox{.}}{2009}]%
        {deng2009imagenet}
\bibfield{author}{\bibinfo{person}{Jia Deng}, \bibinfo{person}{Wei Dong},
  \bibinfo{person}{Richard Socher}, \bibinfo{person}{Li-Jia Li},
  \bibinfo{person}{Kai Li}, {and} \bibinfo{person}{Li Fei-Fei}.}
  \bibinfo{year}{2009}\natexlab{}.
\newblock \showarticletitle{Imagenet: A large-scale hierarchical image
  database}. In \bibinfo{booktitle}{\emph{2009 IEEE conference on computer
  vision and pattern recognition}}.
\newblock


\bibitem[\protect\citeauthoryear{Devlin, Chang, Lee, and Toutanova}{Devlin
  et~al\mbox{.}}{2019}]%
        {devlin2019bert}
\bibfield{author}{\bibinfo{person}{Jacob Devlin}, \bibinfo{person}{Ming-Wei
  Chang}, \bibinfo{person}{Kenton Lee}, {and} \bibinfo{person}{Kristina
  Toutanova}.} \bibinfo{year}{2019}\natexlab{}.
\newblock \showarticletitle{BERT: Pre-training of Deep Bidirectional
  Transformers for Language Understanding}. In \bibinfo{booktitle}{\emph{North
  American Chapter of the Association for Computational Linguistics}}.
\newblock


\bibitem[\protect\citeauthoryear{Fredrikson, Jha, and Ristenpart}{Fredrikson
  et~al\mbox{.}}{2015}]%
        {fredrikson2015model}
\bibfield{author}{\bibinfo{person}{Matt Fredrikson}, \bibinfo{person}{Somesh
  Jha}, {and} \bibinfo{person}{Thomas Ristenpart}.}
  \bibinfo{year}{2015}\natexlab{}.
\newblock \showarticletitle{Model inversion attacks that exploit confidence
  information and basic countermeasures}. In \bibinfo{booktitle}{\emph{The 22nd
  ACM SIGSAC Conference on Computer and Communications Security}}.
\newblock


\bibitem[\protect\citeauthoryear{Fredrikson, Lantz, Jha, Lin, Page, and
  Ristenpart}{Fredrikson et~al\mbox{.}}{2014}]%
        {fredrikson2014privacy}
\bibfield{author}{\bibinfo{person}{Matthew Fredrikson}, \bibinfo{person}{Eric
  Lantz}, \bibinfo{person}{Somesh Jha}, \bibinfo{person}{Simon Lin},
  \bibinfo{person}{David Page}, {and} \bibinfo{person}{Thomas Ristenpart}.}
  \bibinfo{year}{2014}\natexlab{}.
\newblock \showarticletitle{Privacy in pharmacogenetics: An end-to-end case
  study of personalized warfarin dosing}. In \bibinfo{booktitle}{\emph{23rd
  $\{$USENIX$\}$ Security Symposium ($\{$USENIX$\}$ Security 14)}}.
\newblock


\bibitem[\protect\citeauthoryear{Gao, Xu, Wang, Chen, Ranasinghe, and
  Nepal}{Gao et~al\mbox{.}}{2019}]%
        {gao2019strip}
\bibfield{author}{\bibinfo{person}{Yansong Gao}, \bibinfo{person}{Chang Xu},
  \bibinfo{person}{Derui Wang}, \bibinfo{person}{Shiping Chen},
  \bibinfo{person}{Damith~C Ranasinghe}, {and} \bibinfo{person}{Surya Nepal}.}
  \bibinfo{year}{2019}\natexlab{}.
\newblock \showarticletitle{STRIP: A Defence Against Trojan Attacks on Deep
  Neural Networks}.
\newblock \bibinfo{journal}{\emph{arXiv preprint arXiv:1902.06531}}
  (\bibinfo{year}{2019}).
\newblock


\bibitem[\protect\citeauthoryear{Goodfellow, Mirza, Xiao, Courville, and
  Bengio}{Goodfellow et~al\mbox{.}}{2013}]%
        {goodfellow2013empirical}
\bibfield{author}{\bibinfo{person}{Ian~J Goodfellow}, \bibinfo{person}{Mehdi
  Mirza}, \bibinfo{person}{Da Xiao}, \bibinfo{person}{Aaron Courville}, {and}
  \bibinfo{person}{Yoshua Bengio}.} \bibinfo{year}{2013}\natexlab{}.
\newblock \showarticletitle{An empirical investigation of catastrophic
  forgetting in gradient-based neural networks}.
\newblock \bibinfo{journal}{\emph{arXiv preprint arXiv:1312.6211}}
  (\bibinfo{year}{2013}).
\newblock


\bibitem[\protect\citeauthoryear{Goodfellow, Shlens, and Szegedy}{Goodfellow
  et~al\mbox{.}}{2015}]%
        {goodfellow2014explaining}
\bibfield{author}{\bibinfo{person}{Ian~J Goodfellow}, \bibinfo{person}{Jonathon
  Shlens}, {and} \bibinfo{person}{Christian Szegedy}.}
  \bibinfo{year}{2015}\natexlab{}.
\newblock \showarticletitle{Explaining and harnessing adversarial examples}.
\newblock \bibinfo{journal}{\emph{International Conference on Learning
  Representations (ICLR)}} (\bibinfo{year}{2015}).
\newblock


\bibitem[\protect\citeauthoryear{Grandvalet and Bengio}{Grandvalet and
  Bengio}{2005}]%
        {grandvalet2005semi}
\bibfield{author}{\bibinfo{person}{Yves Grandvalet} {and}
  \bibinfo{person}{Yoshua Bengio}.} \bibinfo{year}{2005}\natexlab{}.
\newblock \showarticletitle{Semi-supervised learning by entropy minimization}.
  In \bibinfo{booktitle}{\emph{Advances in neural information processing
  systems}}.
\newblock


\bibitem[\protect\citeauthoryear{Gu, Dolan-Gavitt, and Garg}{Gu
  et~al\mbox{.}}{2017}]%
        {gu2017badnets}
\bibfield{author}{\bibinfo{person}{Tianyu Gu}, \bibinfo{person}{Brendan
  Dolan-Gavitt}, {and} \bibinfo{person}{Siddharth Garg}.}
  \bibinfo{year}{2017}\natexlab{}.
\newblock \showarticletitle{BadNets: Identifying Vulnerabilities in the Machine
  Learning Model Supply Chain}.
\newblock \bibinfo{journal}{\emph{arXiv preprint arXiv:1708.06733}}
  (\bibinfo{year}{2017}).
\newblock


\bibitem[\protect\citeauthoryear{Guo, Wang, Xing, Du, and Song}{Guo
  et~al\mbox{.}}{2019}]%
        {guo2019tabor}
\bibfield{author}{\bibinfo{person}{Wenbo Guo}, \bibinfo{person}{Lun Wang},
  \bibinfo{person}{Xinyu Xing}, \bibinfo{person}{Min Du}, {and}
  \bibinfo{person}{Dawn Song}.} \bibinfo{year}{2019}\natexlab{}.
\newblock \showarticletitle{TABOR: A Highly Accurate Approach to Inspecting and
  Restoring Trojan Backdoors in AI Systems}.
\newblock \bibinfo{journal}{\emph{arXiv preprint arXiv:1908.01763}}
  (\bibinfo{year}{2019}).
\newblock


\bibitem[\protect\citeauthoryear{Hayes, Melis, Danezis, and
  De~Cristofaro}{Hayes et~al\mbox{.}}{2019}]%
        {hayes2019logan}
\bibfield{author}{\bibinfo{person}{Jamie Hayes}, \bibinfo{person}{Luca Melis},
  \bibinfo{person}{George Danezis}, {and} \bibinfo{person}{Emiliano
  De~Cristofaro}.} \bibinfo{year}{2019}\natexlab{}.
\newblock \showarticletitle{LOGAN: Membership inference attacks against
  generative models}.
\newblock \bibinfo{journal}{\emph{Proceedings on Privacy Enhancing
  Technologies}} (\bibinfo{year}{2019}).
\newblock


\bibitem[\protect\citeauthoryear{He, Zhang, Ren, and Sun}{He
  et~al\mbox{.}}{2016}]%
        {he2016deep}
\bibfield{author}{\bibinfo{person}{Kaiming He}, \bibinfo{person}{Xiangyu
  Zhang}, \bibinfo{person}{Shaoqing Ren}, {and} \bibinfo{person}{Jian Sun}.}
  \bibinfo{year}{2016}\natexlab{}.
\newblock \showarticletitle{Deep residual learning for image recognition}. In
  \bibinfo{booktitle}{\emph{IEEE conference on computer vision and pattern
  recognition}}.
\newblock


\bibitem[\protect\citeauthoryear{Hinton, Vinyals, and Dean}{Hinton
  et~al\mbox{.}}{2015}]%
        {hinton2015distilling}
\bibfield{author}{\bibinfo{person}{Geoffrey Hinton}, \bibinfo{person}{Oriol
  Vinyals}, {and} \bibinfo{person}{Jeff Dean}.}
  \bibinfo{year}{2015}\natexlab{}.
\newblock \showarticletitle{Distilling the knowledge in a neural network}.
\newblock \bibinfo{journal}{\emph{arXiv preprint arXiv:1503.02531}}
  (\bibinfo{year}{2015}).
\newblock


\bibitem[\protect\citeauthoryear{Hitaj and Mancini}{Hitaj and Mancini}{2018}]%
        {hitaj2018have}
\bibfield{author}{\bibinfo{person}{Dorjan Hitaj} {and} \bibinfo{person}{Luigi~V
  Mancini}.} \bibinfo{year}{2018}\natexlab{}.
\newblock \showarticletitle{Have you stolen my model? evasion attacks against
  deep neural network watermarking techniques}.
\newblock \bibinfo{journal}{\emph{arXiv preprint arXiv:1809.00615}}
  (\bibinfo{year}{2018}).
\newblock


\bibitem[\protect\citeauthoryear{Kemker, McClure, Abitino, Hayes, and
  Kanan}{Kemker et~al\mbox{.}}{2018}]%
        {kemker2018measuring}
\bibfield{author}{\bibinfo{person}{Ronald Kemker}, \bibinfo{person}{Marc
  McClure}, \bibinfo{person}{Angelina Abitino}, \bibinfo{person}{Tyler~L
  Hayes}, {and} \bibinfo{person}{Christopher Kanan}.}
  \bibinfo{year}{2018}\natexlab{}.
\newblock \showarticletitle{Measuring catastrophic forgetting in neural
  networks}. In \bibinfo{booktitle}{\emph{AAAI conference on artificial
  intelligence}}.
\newblock


\bibitem[\protect\citeauthoryear{Kirkpatrick, Pascanu, Rabinowitz, Veness,
  Desjardins, Rusu, Milan, Quan, Ramalho, Grabska-Barwinska,
  et~al\mbox{.}}{Kirkpatrick et~al\mbox{.}}{2017}]%
        {kirkpatrick2017overcoming}
\bibfield{author}{\bibinfo{person}{James Kirkpatrick}, \bibinfo{person}{Razvan
  Pascanu}, \bibinfo{person}{Neil Rabinowitz}, \bibinfo{person}{Joel Veness},
  \bibinfo{person}{Guillaume Desjardins}, \bibinfo{person}{Andrei~A Rusu},
  \bibinfo{person}{Kieran Milan}, \bibinfo{person}{John Quan},
  \bibinfo{person}{Tiago Ramalho}, \bibinfo{person}{Agnieszka
  Grabska-Barwinska}, {et~al\mbox{.}}} \bibinfo{year}{2017}\natexlab{}.
\newblock \showarticletitle{Overcoming catastrophic forgetting in neural
  networks}.
\newblock \bibinfo{journal}{\emph{Proceedings of the national academy of
  sciences}} (\bibinfo{year}{2017}).
\newblock


\bibitem[\protect\citeauthoryear{Koh and Liang}{Koh and Liang}{2017}]%
        {koh2017understanding}
\bibfield{author}{\bibinfo{person}{Pang~Wei Koh} {and} \bibinfo{person}{Percy
  Liang}.} \bibinfo{year}{2017}\natexlab{}.
\newblock \showarticletitle{Understanding Black-box Predictions via Influence
  Functions}. In \bibinfo{booktitle}{\emph{International Conference on Machine
  Learning}}. \bibinfo{pages}{1885--1894}.
\newblock


\bibitem[\protect\citeauthoryear{Krizhevsky et~al\mbox{.}}{Krizhevsky
  et~al\mbox{.}}{2009}]%
        {krizhevsky2009learning}
\bibfield{author}{\bibinfo{person}{Alex Krizhevsky} {et~al\mbox{.}}}
  \bibinfo{year}{2009}\natexlab{}.
\newblock \bibinfo{booktitle}{\emph{Learning multiple layers of features from
  tiny images}}.
\newblock \bibinfo{type}{{T}echnical {R}eport}.
  \bibinfo{institution}{Citeseer}.
\newblock


\bibitem[\protect\citeauthoryear{Li, Wang, Singh, and Vorobeychik}{Li
  et~al\mbox{.}}{2016}]%
        {li2016data}
\bibfield{author}{\bibinfo{person}{Bo Li}, \bibinfo{person}{Yining Wang},
  \bibinfo{person}{Aarti Singh}, {and} \bibinfo{person}{Yevgeniy Vorobeychik}.}
  \bibinfo{year}{2016}\natexlab{}.
\newblock \showarticletitle{Data poisoning attacks on factorization-based
  collaborative filtering}. In \bibinfo{booktitle}{\emph{Advances in neural
  information processing systems}}.
\newblock


\bibitem[\protect\citeauthoryear{Liu, Dolan-Gavitt, and Garg}{Liu
  et~al\mbox{.}}{2018}]%
        {liu2018fine}
\bibfield{author}{\bibinfo{person}{Kang Liu}, \bibinfo{person}{Brendan
  Dolan-Gavitt}, {and} \bibinfo{person}{Siddharth Garg}.}
  \bibinfo{year}{2018}\natexlab{}.
\newblock \showarticletitle{Fine-pruning: Defending against backdooring attacks
  on deep neural networks}. In \bibinfo{booktitle}{\emph{International
  Symposium on Research in Attacks, Intrusions, and Defenses}}.
\newblock


\bibitem[\protect\citeauthoryear{Liu, Ma, Aafer, Lee, Zhai, Wang, and
  Zhang}{Liu et~al\mbox{.}}{2017a}]%
        {liu2017trojaning}
\bibfield{author}{\bibinfo{person}{Yingqi Liu}, \bibinfo{person}{Shiqing Ma},
  \bibinfo{person}{Yousra Aafer}, \bibinfo{person}{Wen-Chuan Lee},
  \bibinfo{person}{Juan Zhai}, \bibinfo{person}{Weihang Wang}, {and}
  \bibinfo{person}{Xiangyu Zhang}.} \bibinfo{year}{2017}\natexlab{a}.
\newblock \showarticletitle{Trojaning Attack on Neural Networks}. In
  \bibinfo{booktitle}{\emph{Network and Distributed System Security Symposium
  (NDSS)}}.
\newblock


\bibitem[\protect\citeauthoryear{Liu, Xie, and Srivastava}{Liu
  et~al\mbox{.}}{2017b}]%
        {liu2017neural}
\bibfield{author}{\bibinfo{person}{Yuntao Liu}, \bibinfo{person}{Yang Xie},
  {and} \bibinfo{person}{Ankur Srivastava}.} \bibinfo{year}{2017}\natexlab{b}.
\newblock \showarticletitle{Neural Trojans}. In \bibinfo{booktitle}{\emph{The
  35th IEEE International Conference on Computer Design}}.
\newblock


\bibitem[\protect\citeauthoryear{Lopez-Paz and Ranzato}{Lopez-Paz and
  Ranzato}{2017}]%
        {lopez2017gradient}
\bibfield{author}{\bibinfo{person}{David Lopez-Paz} {and}
  \bibinfo{person}{Marc'Aurelio Ranzato}.} \bibinfo{year}{2017}\natexlab{}.
\newblock \showarticletitle{Gradient episodic memory for continual learning}.
  In \bibinfo{booktitle}{\emph{Advances in Neural Information Processing
  Systems}}.
\newblock


\bibitem[\protect\citeauthoryear{Merrer, Perez, and Tr{\'e}dan}{Merrer
  et~al\mbox{.}}{2019}]%
        {merrer2017adversarial}
\bibfield{author}{\bibinfo{person}{Erwan~Le Merrer}, \bibinfo{person}{Patrick
  Perez}, {and} \bibinfo{person}{Gilles Tr{\'e}dan}.}
  \bibinfo{year}{2019}\natexlab{}.
\newblock \showarticletitle{Adversarial frontier stitching for remote neural
  network watermarking}.
\newblock \bibinfo{journal}{\emph{Journal of Neural Computing and
  Applications}} (\bibinfo{year}{2019}).
\newblock


\bibitem[\protect\citeauthoryear{Miyato, Maeda, Koyama, and Ishii}{Miyato
  et~al\mbox{.}}{2018}]%
        {miyato2018virtual}
\bibfield{author}{\bibinfo{person}{Takeru Miyato}, \bibinfo{person}{Shin-ichi
  Maeda}, \bibinfo{person}{Masanori Koyama}, {and} \bibinfo{person}{Shin
  Ishii}.} \bibinfo{year}{2018}\natexlab{}.
\newblock \showarticletitle{Virtual adversarial training: a regularization
  method for supervised and semi-supervised learning}.
\newblock \bibinfo{journal}{\emph{IEEE transactions on pattern analysis and
  machine intelligence}} (\bibinfo{year}{2018}).
\newblock


\bibitem[\protect\citeauthoryear{Mu{\~n}oz-Gonz{\'a}lez, Biggio, Demontis,
  Paudice, Wongrassamee, Lupu, and Roli}{Mu{\~n}oz-Gonz{\'a}lez
  et~al\mbox{.}}{2017}]%
        {munoz2017towards}
\bibfield{author}{\bibinfo{person}{Luis Mu{\~n}oz-Gonz{\'a}lez},
  \bibinfo{person}{Battista Biggio}, \bibinfo{person}{Ambra Demontis},
  \bibinfo{person}{Andrea Paudice}, \bibinfo{person}{Vasin Wongrassamee},
  \bibinfo{person}{Emil~C Lupu}, {and} \bibinfo{person}{Fabio Roli}.}
  \bibinfo{year}{2017}\natexlab{}.
\newblock \showarticletitle{Towards poisoning of deep learning algorithms with
  back-gradient optimization}. In \bibinfo{booktitle}{\emph{10th ACM Workshop
  on Artificial Intelligence and Security}}.
\newblock


\bibitem[\protect\citeauthoryear{Namba and Sakuma}{Namba and Sakuma}{2019}]%
        {namba2019robust}
\bibfield{author}{\bibinfo{person}{Ryota Namba} {and} \bibinfo{person}{Jun
  Sakuma}.} \bibinfo{year}{2019}\natexlab{}.
\newblock \showarticletitle{Robust Watermarking of Neural Network with
  Exponential Weighting}. In \bibinfo{booktitle}{\emph{2019 {ACM} Asia
  Conference on Computer and Communications Security}}.
\newblock


\bibitem[\protect\citeauthoryear{Nelson, Barreno, Chi, Joseph, Rubinstein,
  Saini, Sutton, Tygar, and Xia}{Nelson et~al\mbox{.}}{2008}]%
        {nelson2008exploiting}
\bibfield{author}{\bibinfo{person}{Blaine Nelson}, \bibinfo{person}{Marco
  Barreno}, \bibinfo{person}{Fuching~Jack Chi}, \bibinfo{person}{Anthony~D
  Joseph}, \bibinfo{person}{Benjamin~IP Rubinstein}, \bibinfo{person}{Udam
  Saini}, \bibinfo{person}{Charles~A Sutton}, \bibinfo{person}{J~Doug Tygar},
  {and} \bibinfo{person}{Kai Xia}.} \bibinfo{year}{2008}\natexlab{}.
\newblock \showarticletitle{Exploiting Machine Learning to Subvert Your Spam
  Filter}.
\newblock \bibinfo{journal}{\emph{LEET}} (\bibinfo{year}{2008}).
\newblock


\bibitem[\protect\citeauthoryear{Papernot, McDaniel, Wu, Jha, and
  Swami}{Papernot et~al\mbox{.}}{2016}]%
        {papernot2016distillation}
\bibfield{author}{\bibinfo{person}{Nicolas Papernot}, \bibinfo{person}{Patrick
  McDaniel}, \bibinfo{person}{Xi Wu}, \bibinfo{person}{Somesh Jha}, {and}
  \bibinfo{person}{Ananthram Swami}.} \bibinfo{year}{2016}\natexlab{}.
\newblock \showarticletitle{Distillation as a defense to adversarial
  perturbations against deep neural networks}. In
  \bibinfo{booktitle}{\emph{2016 IEEE Symposium on Security and Privacy (SP)}}.
\newblock


\bibitem[\protect\citeauthoryear{Rouhani, Chen, and Koushanfar}{Rouhani
  et~al\mbox{.}}{2018}]%
        {rouhani2018deepsigns}
\bibfield{author}{\bibinfo{person}{Bita~Darvish Rouhani},
  \bibinfo{person}{Huili Chen}, {and} \bibinfo{person}{Farinaz Koushanfar}.}
  \bibinfo{year}{2018}\natexlab{}.
\newblock \showarticletitle{Deepsigns: A generic watermarking framework for ip
  protection of deep learning models}.
\newblock \bibinfo{journal}{\emph{arXiv preprint arXiv:1804.00750}}
  (\bibinfo{year}{2018}).
\newblock


\bibitem[\protect\citeauthoryear{Shafahi, Huang, Najibi, Suciu, Studer,
  Dumitras, and Goldstein}{Shafahi et~al\mbox{.}}{2018}]%
        {shafahi2018poison}
\bibfield{author}{\bibinfo{person}{Ali Shafahi}, \bibinfo{person}{W~Ronny
  Huang}, \bibinfo{person}{Mahyar Najibi}, \bibinfo{person}{Octavian Suciu},
  \bibinfo{person}{Christoph Studer}, \bibinfo{person}{Tudor Dumitras}, {and}
  \bibinfo{person}{Tom Goldstein}.} \bibinfo{year}{2018}\natexlab{}.
\newblock \showarticletitle{Poison frogs! targeted clean-label poisoning
  attacks on neural networks}.
\newblock \bibinfo{journal}{\emph{Advances in Neural Information Processing
  Systems}} (\bibinfo{year}{2018}).
\newblock


\bibitem[\protect\citeauthoryear{Shin, Lee, Kim, and Kim}{Shin
  et~al\mbox{.}}{2017}]%
        {shin2017continual}
\bibfield{author}{\bibinfo{person}{Hanul Shin}, \bibinfo{person}{Jung~Kwon
  Lee}, \bibinfo{person}{Jaehong Kim}, {and} \bibinfo{person}{Jiwon Kim}.}
  \bibinfo{year}{2017}\natexlab{}.
\newblock \showarticletitle{Continual learning with deep generative replay}. In
  \bibinfo{booktitle}{\emph{Advances in Neural Information Processing
  Systems}}.
\newblock


\bibitem[\protect\citeauthoryear{Shokri, Stronati, Song, and Shmatikov}{Shokri
  et~al\mbox{.}}{2017}]%
        {shokri2017membership}
\bibfield{author}{\bibinfo{person}{Reza Shokri}, \bibinfo{person}{Marco
  Stronati}, \bibinfo{person}{Congzheng Song}, {and} \bibinfo{person}{Vitaly
  Shmatikov}.} \bibinfo{year}{2017}\natexlab{}.
\newblock \showarticletitle{Membership inference attacks against machine
  learning models}. In \bibinfo{booktitle}{\emph{2017 IEEE Symposium on
  Security and Privacy (SP)}}.
\newblock


\bibitem[\protect\citeauthoryear{Simonyan and Zisserman}{Simonyan and
  Zisserman}{2015}]%
        {simonyan2014very}
\bibfield{author}{\bibinfo{person}{Karen Simonyan} {and}
  \bibinfo{person}{Andrew Zisserman}.} \bibinfo{year}{2015}\natexlab{}.
\newblock \showarticletitle{Very deep convolutional networks for large-scale
  image recognition}.
\newblock \bibinfo{journal}{\emph{International Conference on Learning
  Representations (ICLR)}} (\bibinfo{year}{2015}).
\newblock


\bibitem[\protect\citeauthoryear{Szegedy, Zaremba, Sutskever, Bruna, Erhan,
  Goodfellow, and Fergus}{Szegedy et~al\mbox{.}}{2013}]%
        {szegedy2013intriguing}
\bibfield{author}{\bibinfo{person}{Christian Szegedy},
  \bibinfo{person}{Wojciech Zaremba}, \bibinfo{person}{Ilya Sutskever},
  \bibinfo{person}{Joan Bruna}, \bibinfo{person}{Dumitru Erhan},
  \bibinfo{person}{Ian Goodfellow}, {and} \bibinfo{person}{Rob Fergus}.}
  \bibinfo{year}{2013}\natexlab{}.
\newblock \showarticletitle{Intriguing properties of neural networks}.
\newblock \bibinfo{journal}{\emph{arXiv preprint arXiv:1312.6199}}
  (\bibinfo{year}{2013}).
\newblock


\bibitem[\protect\citeauthoryear{Tran, Li, and Madry}{Tran
  et~al\mbox{.}}{2018}]%
        {tran2018spectral}
\bibfield{author}{\bibinfo{person}{Brandon Tran}, \bibinfo{person}{Jerry Li},
  {and} \bibinfo{person}{Aleksander Madry}.} \bibinfo{year}{2018}\natexlab{}.
\newblock \showarticletitle{Spectral signatures in backdoor attacks}. In
  \bibinfo{booktitle}{\emph{Advances in Neural Information Processing
  Systems}}.
\newblock


\bibitem[\protect\citeauthoryear{Uchida, Nagai, Sakazawa, and Satoh}{Uchida
  et~al\mbox{.}}{2017}]%
        {uchida2017embedding}
\bibfield{author}{\bibinfo{person}{Yusuke Uchida}, \bibinfo{person}{Yuki
  Nagai}, \bibinfo{person}{Shigeyuki Sakazawa}, {and}
  \bibinfo{person}{Shin'ichi Satoh}.} \bibinfo{year}{2017}\natexlab{}.
\newblock \showarticletitle{Embedding watermarks into deep neural networks}. In
  \bibinfo{booktitle}{\emph{ACM International Conference on Multimedia
  Retrieval}}.
\newblock


\bibitem[\protect\citeauthoryear{Wang, Yao, Shan, Li, Viswanath, Zheng, and
  Zhao}{Wang et~al\mbox{.}}{2019}]%
        {wang2019neural}
\bibfield{author}{\bibinfo{person}{Bolun Wang}, \bibinfo{person}{Yuanshun Yao},
  \bibinfo{person}{Shawn Shan}, \bibinfo{person}{Huiying Li},
  \bibinfo{person}{Bimal Viswanath}, \bibinfo{person}{Haitao Zheng}, {and}
  \bibinfo{person}{Ben~Y Zhao}.} \bibinfo{year}{2019}\natexlab{}.
\newblock \showarticletitle{Neural cleanse: Identifying and mitigating backdoor
  attacks in neural networks}. In \bibinfo{booktitle}{\emph{IEEE Symposium on
  Security and Privacy}}.
\newblock


\bibitem[\protect\citeauthoryear{Yang, Dang, and Chang}{Yang
  et~al\mbox{.}}{2019}]%
        {yang2019effectiveness}
\bibfield{author}{\bibinfo{person}{Ziqi Yang}, \bibinfo{person}{Hung Dang},
  {and} \bibinfo{person}{Ee-Chien Chang}.} \bibinfo{year}{2019}\natexlab{}.
\newblock \showarticletitle{Effectiveness of Distillation Attack and
  Countermeasure on Neural Network Watermarking}.
\newblock \bibinfo{journal}{\emph{arXiv preprint arXiv:1906.06046}}
  (\bibinfo{year}{2019}).
\newblock


\bibitem[\protect\citeauthoryear{Yosinski, Clune, Bengio, and Lipson}{Yosinski
  et~al\mbox{.}}{2014}]%
        {yosinski2014transferable}
\bibfield{author}{\bibinfo{person}{Jason Yosinski}, \bibinfo{person}{Jeff
  Clune}, \bibinfo{person}{Yoshua Bengio}, {and} \bibinfo{person}{Hod Lipson}.}
  \bibinfo{year}{2014}\natexlab{}.
\newblock \showarticletitle{How transferable are features in deep neural
  networks?}. In \bibinfo{booktitle}{\emph{Advances in neural information
  processing systems}}.
\newblock


\bibitem[\protect\citeauthoryear{Zenke, Poole, and Ganguli}{Zenke
  et~al\mbox{.}}{2017}]%
        {zenke2017continual}
\bibfield{author}{\bibinfo{person}{Friedemann Zenke}, \bibinfo{person}{Ben
  Poole}, {and} \bibinfo{person}{Surya Ganguli}.}
  \bibinfo{year}{2017}\natexlab{}.
\newblock \showarticletitle{Continual learning through synaptic intelligence}.
  In \bibinfo{booktitle}{\emph{International Conference on Machine Learning}}.
\newblock


\bibitem[\protect\citeauthoryear{Zhang, Gu, Jang, Wu, Stoecklin, Huang, and
  Molloy}{Zhang et~al\mbox{.}}{2018}]%
        {zhang2018protecting}
\bibfield{author}{\bibinfo{person}{Jialong Zhang}, \bibinfo{person}{Zhongshu
  Gu}, \bibinfo{person}{Jiyong Jang}, \bibinfo{person}{Hui Wu},
  \bibinfo{person}{Marc~Ph Stoecklin}, \bibinfo{person}{Heqing Huang}, {and}
  \bibinfo{person}{Ian Molloy}.} \bibinfo{year}{2018}\natexlab{}.
\newblock \showarticletitle{Protecting intellectual property of deep neural
  networks with watermarking}. In \bibinfo{booktitle}{\emph{Asia Conference on
  Computer and Communications Security}}.
\newblock


\end{thebibliography}

\appendix
\section{More Discussion on Experimental Details}
\label{app:exp}

\begin{figure}[tbp]
    \centering
    \includegraphics[width=\linewidth]{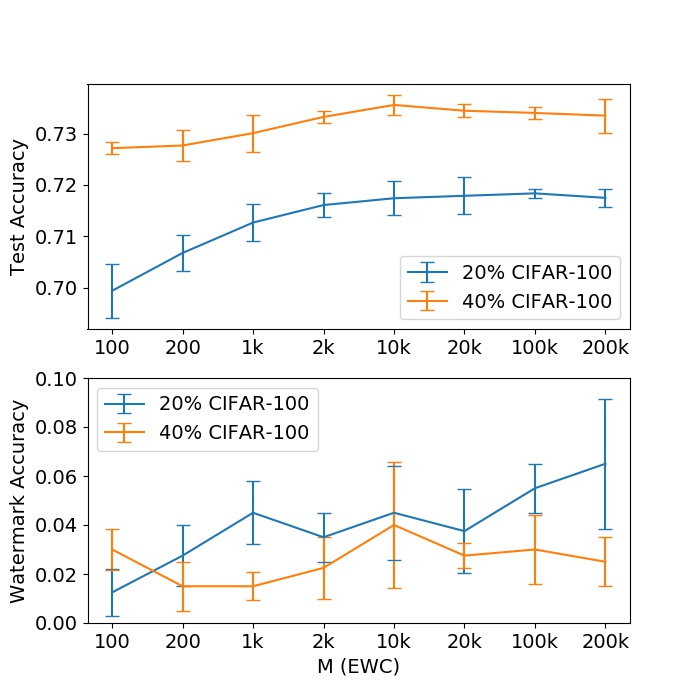}
    \caption{Results with different $M$ for EWC, where we perform 4 runs for each $M$, and plot the mean and standard deviation. The pre-trained models are embedded with pattern-based watermarks, and fine-tuned with partial CIFAR-100.}
    \label{fig:ewc-m}
\end{figure}

\begin{figure}[tbp]
    \centering
    \includegraphics[width=\linewidth]{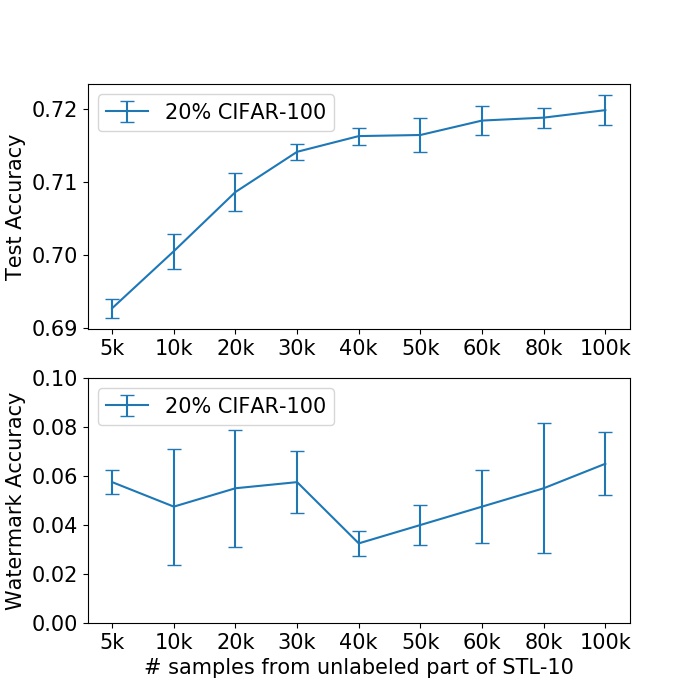}
    \caption{Results with different numbers of unlabeled samples for AU, where we perform 4 runs for every setting and plot the mean and standard deviation. The pre-trained models are embedded with pattern-based watermarks and fine-tuned with partial CIFAR-100. The unlabeled samples for augmentation are drawn from the unlabeled part of STL-10.}
    \label{fig:au-sample-size}
\end{figure}

\begin{figure}[tbp]
    \centering
    \includegraphics[width=\linewidth]{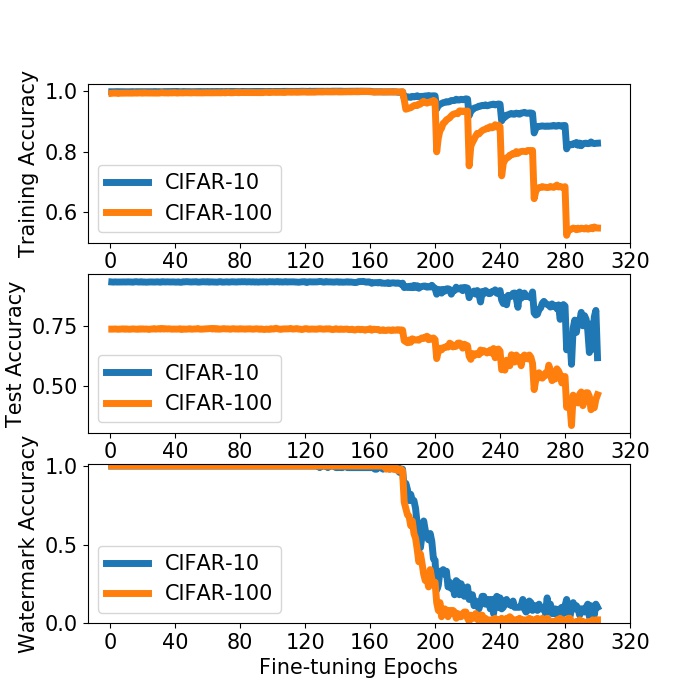}
    \caption{Training curves to illustrate the effect of learning rate during the fine-tuning stage. The configuration is mostly the same as Figure~\ref{fig:double-lr-partial02-20}, except that the model is fine-tuned on the entire training set.}
    \label{fig:double-lr-20}
\end{figure}

\begin{table*}[t]
    \centering
    \begin{tabular}{|c|c|ccc|}
    \toprule
        Dataset & Scheme & Initial learning rate & $\lambda$(EWC) & m(AU) \\
    \midrule
        \multirow{4}{*}{CIFAR-10} & Pattern & 0.03 & 150 & 50
        \\
        & OOD & $[0.05, 0.15]$& $10$ & $50$
        \\
        & EW & $0.03$ & $20$ & $50$
        \\
        & AFS &$[0.01, 0.1]$ & $3$ & $[5, 50]$
        \\
    \midrule
        \multirow{4}{*}{CIFAR-100} & Pattern & $[0.03, 0.1]$ & $20$ & $50$
        \\
        & OOD & $[0.03, 0.1]$ & $200$ & $50$
        \\
        & EW & $[0.04, 0.05]$&  $[2, 5]$& $50$
        \\
        & AFS & $[0.015, 0.07]$ & $[25, 30]$ & $[10, 50]$
        \\
    \midrule
        \multirow{4}{*}{CIFAR-10$\to$ STL-10} & Pattern & $[0.03, 0.05]$ & $10$ & $50$
        \\
        & OOD & $[0.04, 0.15]$ & $10$ & $50$
        \\
        & EW & $[0.03, 0.05]$ & $200$ & $50$
        \\
        & AFS & $[0.02, 0.05]$ & $200$ & $50$
        \\
    \midrule
        \multirow{4}{*}{ImageNet32} & Pattern & $[0.004, 0.04]$ & $[800, 1200]$ & $50$
        \\
        & OOD & $[0.005, 0.05]$ & $[30, 100]$ & $50$
        \\
        & EW & $[0.003, 0.1]$ & $[10^4, 2\times 10^4]$ & $50$
        \\
        & AFS & $[0.006, 0.03]$ & $[3, 50]$ & $[30, 50]$
        \\
    \midrule
        \multirow{4}{*}{ImageNet32$\to$ STL-10} & Pattern & $[0.015, 0.02]$ & $[1000, 1100]$ & $50$
        \\
        & OOD & $[0.01, 0.015]$ & $[50, 100]$ & $50$
        \\
        & EW & $[0.007, 0.03]$ & $[1.2 \times 10^4, 1.5 \times 10^4]$ & $50$
        \\
        & AFS & $[0.003, 0.008]$ & $[200, 500]$ & $50$
        \\
    \bottomrule
    \end{tabular}
    \caption{Ranges of the best hyper-parameter configuration for all watermark removal results. $\lambda$ denotes the coefficient in EWC and $m$ is the number of unlabeled samples added to a training batch with AU. }
    \label{tab:hyperparameters}
\end{table*}

Our implementation is in PyTorch~\footnote{The implementation is mainly adapted from~\url{https://github.com/adiyoss/WatermarkNN}, the code repo of~\cite{adi2018turning}.}. For each watermarking scheme in our evaluation, we present the best hyper-parameter configurations in Table~\ref{tab:hyperparameters}. Note that the adversary can always make the worst assumption about the strength of the watermark scheme, and conservatively set the initial learning rate to ensure that the watermarks are removed. In our evaluation, we observe that setting an initial learning rate to be $0.05$ works relatively well for all settings. Other hyper-parameters are set to maximize the test accuracy after watermark removal.

\section{Discussion of sample efficiency}
\label{app:eval-sample-efficiency}

In the following, we provide some discussion about the sample efficiency of EWC and AU components in {\ours}.

\paragraph{The number of samples $M$ for EWC.} For EWC component, we investigate how the number of samples $M$ drawn for Fisher information approximation affects the performance, and present the results in Figure~\ref{fig:ewc-m}. Specifically, we evaluate on CIFAR-100, and embed the pre-trained models with pattern-based watermarks. We observe that with only $M=100$ samples, EWC is already able to increase the test accuracy around $1\%$ over the basic fine-tuning, demonstrating its effectiveness of preserving the test performance. Setting a higher $M$ may further improve the results, but it could also introduce a higher computation overhead without significant performance gain when $M$ becomes too large. Therefore, we set the default value of $M$ based on such a trade-off.

\paragraph{The number of unlabeled samples for AU.} For AU component, we demonstrate the results of varying the number of unlabeled samples for fine-tuning in Figure~\ref{fig:au-sample-size}. We also evaluate on CIFAR-100 with the pre-trained model using the pattern-based watermarking scheme, and the unlabeled samples are drawn from the unlabeled part of STL-10. Despite the large difference of dataset distribution between STL-10 and CIFAR-100, augmenting with $5K$ unlabeled samples already enables a considerable performance gain, and the test accuracy continues to increase with more unlabeled samples, suggesting the promise of leveraging unlabeled data for watermark removal, which is typically much easier to collect than in-distribution labeled data.

\section{More Discussion on Pruning}
\label{app:pruning}
Previous work has studied the effectiveness of pruning-based approaches for watermark removal, and found that such techniques are largely ineffective~\cite{zhang2018protecting,liu2018fine,namba2019robust}. In our evaluation, we compare with the pruning method studied in~\cite{liu2018fine}, where we follow their setup to prune the neurons of the last convolutional layer in the increasing order of the magnitude of their activations on the validation set.

\begin{figure*}[t]
    \centering
    \begin{subfigure}[t]{0.48\linewidth}
    \includegraphics[width=\linewidth]{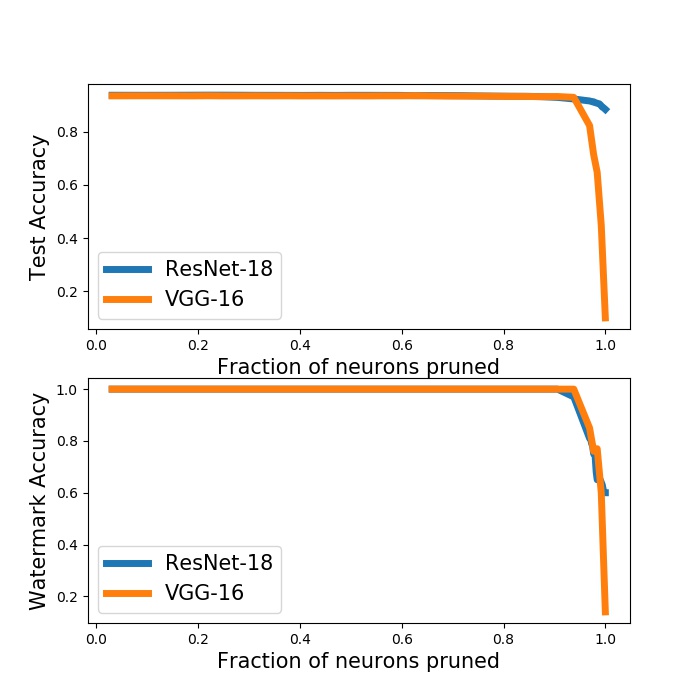}
    \caption{}
    \label{}
    \end{subfigure}
    \begin{subfigure}[t]{0.48\linewidth}
    \includegraphics[width=\linewidth]{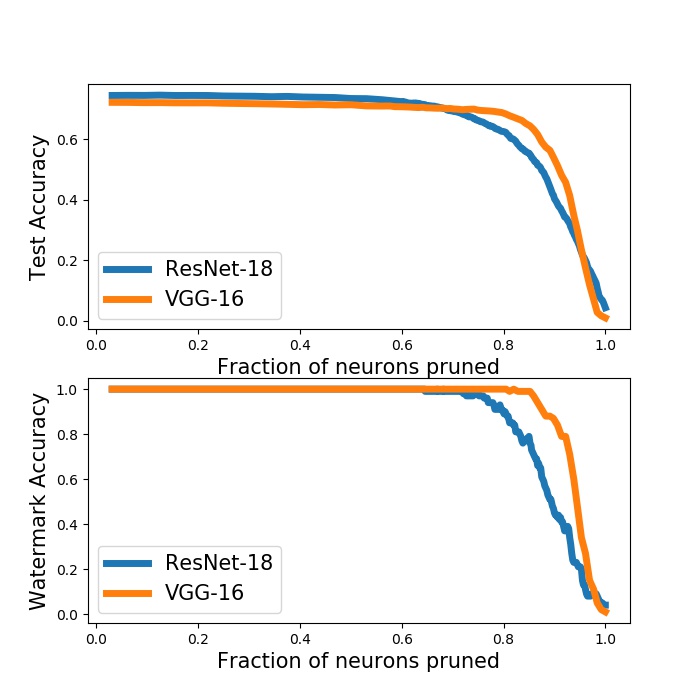}
    \caption{}
    \label{}
    \end{subfigure}
    \caption{Curves to illustrate the effect of neuron pruning. The corresponding pre-trained models are embedded with OOD watermarks. (a) CIFAR-10; (b) CIFAR-100.}
    \label{fig:prune}
\end{figure*}

Figure~\ref{fig:prune} presents the curves of the model accuracy with different pruning rates. Note that due to the skip connections introduced in ResNet architecture, the model accuracy may not be low even if the pruning rate is close to 1. Therefore, we also evaluate VGG-16~\cite{simonyan2014very}, another neural network architecture that is capable of achieving the same level of performance on both CIFAR-10 and CIFAR-100. For both models, we observe that the watermark accuracy is tightly associated with the test accuracy, which makes it hard to find a sweet spot of the pruning rate so that the test performance is preserved while the watermarks are removed.

In particular, as shown in Table~\ref{tab:res-finepruning}, using the pruning approach, when the test accuracy degrades to $90.72\%$ on CIFAR-10, the watermark accuracy is still $65\%$; on the other hand, using {\ours} with AU, without any in-distribution labeled data, the fine-tuned model achieves the same level of performance as the pruning method with the watermarks removed. The gap on CIFAR-100 is more significant: {\ours} is able to achieve an accuracy of $66.79\%$, but the test accuracy of the pruned model already decreases to $53.34\%$ with $71\%$ watermarks still retained. We have also tried other pruning approaches, but none of them works considerably better, which shows that {\ours} is more suitable for watermark removal.

\section{More Discussion on Fine-pruning}
\label{app:fine-pruning}

\begin{table*}[t]
    \centering
    \begin{tabular}{|c|c|c|c|ccccc|}
\hline
\multirow{2}{*}{Dataset} & \multirow{2}{*}{Model} & \multirow{2}{*}{Pruning} & \multirow{2}{*}{Before fine-tuning} & \multicolumn{5}{c|}{Percentage}\\
& & & & 20\% & 30\% & 40\% & 50\% & 80\%  \\
\hline
\multirow{4}{*}{CIFAR-10} & \multirow{2}{*}{ResNet-18} & $\circ$ & $90.72\%(65\%)$ & $91.10\%$ & $92.05\%$ & $92.72\%$ & $93.25\%$ & $94.20\%$\\
  & & $\times$ & $93.73\%(100\%)$ & $90.82\%$ & $92.27\%$ & $92.78\%$ & $93.44\%$ & $94.03\%$ \\
  \cline{2-9}
  & \multirow{2}{*}{VGG-16} & $\circ$ & $64.69\%(77\%)$ & $90.21\%$ & $91.44\%$ & $92.00\%$ & $92.81\%$ & $93.52\%$ \\
  & &$\times$ & $93.48\%(100\%)$ & $89.94\%$ & $91.53\%$ & $92.59\%$ & $92.69\%$ & $93.35\%$ \\
  \hline
\multirow{4}{*}{CIFAR-100} & \multirow{2}{*}{ResNet-18} &$\circ$ & $53.34\%(71\%)$ & $67.34\%$ & $70.25\%$& $71.42\%$ & $72.80\%$ & $74.05\%$ \\
  & & $\times$ & $74.50\%(100\%)$ & $67.83\%$ & $70.54\%$ & $72.16\%$ & $72.49\%$ & $74.74\%$ \\
  \cline{2-9}
  & \multirow{2}{*}{VGG-16} & $\circ$& $63.26\%(97\%)$ & $62.03\%$ & $65.44\%$ & $67.72\%$ & $68.49\%$ & $70.99\%$ \\
  & & $\times$ &$72.19\%(100\%)$ & $62.80\%$ & $65.65\%$ & $68.11\%$ & $69.47\%$ & $71.38\%$ \\
\hline
    \end{tabular}
    \caption{Comparisons between the basic version of {\ours} and fine-pruning~\cite{liu2018fine}, where $\times$ in the column ``Pruning'' denotes {\ours} without EWC and AU, and $\circ$ denotes fine-pruning. The pre-trained models are embedded with OOD watermarks. For results before fine-tuning, we also present the watermark accuracies in the brackets. In the columns of ``Percentage'', we present the proportion of labeled training set used for fine-tuning. For fine-pruning, the ratios of the pruned neurons from the last convolution layer are $98.4\%$ and $85.9\%$ for CIFAR-10 and CIFAR-100, respectively. Note that we apply the same learning rate schedule for fine-pruning as {\ours}, which is crucial in preserving a good test performance while removing the watermarks.}
    \label{tab:res-finepruning}
\end{table*}

For the implementation of fine-pruning, we set the pruning rates before fine-tuning in the same way as their paper, i.e., keep increasing the pruning rate stepwise, and stop when the degradation of the model performance becomes observable. We apply the same learning rate schedule for fine-pruning as {\ours}, which is crucial in preserving the test performance of the model while removing the watermarks.

Table~\ref{tab:res-finepruning} presents the results on CIFAR-10 and CIFAR-100, comparing the fine-pruning approach to the basic version of {\ours} without EWC and AU, where the pre-trained models are embedded with OOD watermarks. Besides ResNet-18, we also evaluate VGG-16~\cite{simonyan2014very}, another neural network architecture that is capable of achieving the same level of performance on both CIFAR-10 and CIFAR-100. For both datasets and model architectures, we find that the results are roughly similar, suggest that pruning is not necessary with a properly designed learning rate schedule for fine-tuning. In particular, our full {\ours} framework still outperforms the fine-pruning. 
\end{document}